\documentclass[12pt,a4paper]{elsarticle}

\usepackage[colorlinks,citecolor=blue,linktoc=all,linkcolor=cyan]{hyperref}
\usepackage{graphicx}

\usepackage[T1]{fontenc}
\usepackage{mathrsfs}             
\usepackage{slashed}              
\usepackage{amsmath}
\usepackage{amssymb}
\usepackage{amsbsy}  
\usepackage{multirow}
\usepackage{amsfonts} 
\usepackage{tikz}
\usepackage{tikz-3dplot}

\usepackage{wasysym} 
\usepackage{booktabs} 

\numberwithin{equation}{section}
\numberwithin{table}{section}
\numberwithin{figure}{section}
\providecommand{\keywords}[1]{\textbf{Keywords:} #1}
\usepackage{ctex}            
\usepackage{amsmath, amssymb} 
\usepackage{geometry}
\usepackage{graphicx}
\usepackage{subcaption}
\usepackage{hyperref}
\renewcommand{\tablename}{Table}
\renewcommand{\figurename}{Figure}
\usepackage{booktabs}
\usepackage{makecell}
\usepackage{siunitx}
\begin{document}
\begin{frontmatter}

\title{
A High-Precision Numerical Framework for Time-Varying Solar Neutrino Flux with Full Earth Matter Oscillation Corrections for Global Underground Laboratories
}
\author[1,2]{Keyu Han}
\author[3]{Isabella Yin}
\author[3]{Kaoru Yagi}
\author[3]{Kevin Yifan Jiang}
\author[4]{Xiangpan Ji}
\author[1,2]{Shaomin Chen}
\address[1]{Center for High Energy Physics, Tsinghua University, Beijing 100084, China}
\address[2]{Department of Engineering Physics, Tsinghua University, Beijing 10084, China}
\address[3]{Tsinghua International School, Beijing 10084, China}
\address[4]{School of Physics, Nankai University,Tianjin, China, 300071}

\begin{abstract}
Solar neutrinos have been studied for over half a century to test both the Standard Solar Model and the electroweak sector of the Standard Model of particle physics. Contemporary experiments are now entering an era of high-precision measurements, demanding corresponding theoretical predictions with sub-percent accuracy to enable meaningful comparison. In this paper, we identify and analyze the essential physical and computational components required to compute solar neutrino fluxes with high fidelity, and present a unified, computationally efficient framework. This framework incorporates: (i) the time-varying Earth–Sun distance; (ii) Earth matter effects modeled using both one-dimensional (1D) and three-dimensional (3D) Earth electron-density profiles; and (iii) a fast, Strang-splitting–based implementation of the Mikheyev–Smirnov–Wolfenstein (MSW) neutrino propagation formalism, enabling rapid, large-scale scans over neutrino trajectories and energy grids. We deliver site-specific predictions for the China Jinping Underground Laboratory (CJPL) and other underground laboratories actively engaged in solar neutrino programs.
\end{abstract}

\end{frontmatter}
\keywords{Solar neutrinos, neutrino oscillation, matter effect, flux, underground laboratory}

\section{Introduction}

Invariance under the combined operations of charge conjugation (C), parity transformation (P), and time reversal (T), collectively known as CPT symmetry, is a cornerstone principle of quantum field theory~\cite{Blum:2022eol}. Its validity guarantees exact equality between particles and antiparticles in fundamental properties, including mass, spin, lifetime, and magnetic moment. Neutrinos, characterized by their exceedingly small masses, electric neutrality, and exclusive participation in weak interactions and gravity, constitute exceptionally sensitive probes for testing CPT symmetry. In particular, potential new physics beyond the Standard Model, such as quantum gravity effects or interactions with dark matter, could induce CPT violation at extremely low energy scales; such violations may be significantly enhanced over the long baselines traversed by neutrinos in propagation. Neutrino flavor oscillations, governed by parameters including mixing angles and squared-mass differences, are thus a highly sensitive observable for CPT symmetry. A violation of CPT symmetry would manifest as systematic discrepancies between the oscillation parameters measured for neutrinos and antineutrinos. Experimentally, a powerful approach is to conduct complementary studies of solar neutrinos (predominantly $\nu_e$) and reactor antineutrinos ($\bar{\nu}_e$). Although both fall within a similar low-energy regime (below a dozen MeV), their production mechanisms are well understood, and their fluxes are sufficiently large to enable robust, high-statistics comparisons between neutrino and antineutrino behavior, thereby providing a stringent test of CPT conservation. This program complements ongoing long-baseline experiments such as T2K and NO$\nu$A, and will further synergize with next-generation facilities including DUNE and Hyper-Kamiokande~\cite{T2K:2018rhz, NOvA:2024lti, DUNE:2015lol, Hyper-KamiokandeProto-:2015xww}.

In solar neutrino experiments, all major components of the solar neutrino energy spectrum, except for the hydrogen–helium fusion (hep) component, have been precisely measured. To date, the precision achieved in determining the solar neutrino mixing parameters (namely, \(\sin^2\theta_{12}\) and \(\Delta m^2_{21}\)) stands at approximately 4\% and 16\%, respectively~\cite{Super-Kamiokande:2023jbt, SNO:2011hxd}.  

According to the CPT invariance principle, reactor neutrino experiments are expected to yield identical values for these mixing parameters within the framework of the standard three-flavor neutrino oscillation model. Nevertheless, historical discrepancies between solar and reactor neutrino measurements have motivated investigations into physics beyond the Standard Model, including sterile neutrino hypotheses and non-standard neutrino interactions~\cite{KamLAND:2013rgu, Maltoni:2015kca}.  

Recently, the Jiangmen Underground Nuclear Astrophysics (JUNO) experiment has performed high-precision measurements of reactor antineutrino oscillations at a baseline of 55 km, improving the uncertainty on \(\sin^2\theta_{12}\) to 3.3\% and demonstrating consistency with solar neutrino results at the level of one standard deviation~\cite{JUNO:2025gmd}. With its projected ultimate precision on \(\sin^2\theta_{12}\) reaching 1.0\%, JUNO will uniquely enable a stringent test of CPT invariance through a joint analysis of solar and reactor neutrino data.
 
As reactor-based neutrino experiments continually extend their operational and analytical frontiers, enhancing the precision of solar neutrino measurements has emerged as a critical bottleneck for performing robust joint tests of neutrino oscillation parameters. Addressing this challenge necessitates a comprehensive, quantitative assessment of all relevant uncertainty sources. Unlike reactor neutrino experiments, where the baseline distance is fixed, the propagation distance of solar neutrinos varies continuously due to both the eccentricity of Earth’s orbit and its diurnal rotation; this time-dependent baseline must therefore be modeled with high fidelity in data analysis.

Moreover, when solar neutrinos traverse the Earth, particularly during nighttime detection, their flavor evolution is modified by matter effects arising from the Earth's electron density profile. This effect is especially pronounced across regions with steep density gradients, such as the core–mantle boundary, and is governed by the Mikheyev–Smirnov–Wolfenstein (MSW) mechanism~\cite{Wolfenstein:1977ue, Mikheyev:1985zog, Mikheev:1986wj}.

Specifically, solar neutrinos that have oscillated into $\nu_\mu$ or $\nu_\tau$ states, either within the Sun’s interior or during vacuum propagation en route to Earth, may undergo matter-induced reverse conversion upon passage through the Earth, regenerating $\nu_e$ components. This neutrino regeneration effect results in a modest but measurable enhancement in the observed $\nu_e$ event rate during nighttime compared to daytime, a phenomenon known as the day–night asymmetry. This asymmetry has been preliminarily observed by experiments such as Super-Kamiokande. Beyond providing crucial validation of neutrino oscillation models, which include MSW-enhanced flavor conversion, it also opens a novel avenue for probing the Earth's internal density structure using neutrinos as natural tomographic probes~\cite{Xu:2022wcq}.

In this paper, we incorporate the time-varying Earth–Sun distance into the neutrino propagation modeling framework by systematically accounting for its periodic variations, arising from Earth’s orbital eccentricity and rotational motion. Leveraging high-precision astronomical ephemeris data, we compute the instantaneous neutrino propagation path length and consistently include Earth-matter effects. Specifically, we evaluate these matter effects using both one-dimensional (1D) and three-dimensional (3D) electron density models. The 1D model is based on a simplified parametrization of the Preliminary Reference Earth Model (PREM)~\cite{PREM}. In contrast, the 3D model integrates state-of-the-art seismic tomography data to resolve steep electron density gradients, particularly at the mantle–outer core boundary with improved fidelity~\cite{LITHO10, SAW642AN}.

Adopting a three-dimensional Earth model introduces explicit azimuthal dependence in addition to zenith-angle dependence, thereby substantially increasing the number of distinct propagation paths requiring evaluation.  
To address this computational challenge, we employ an unsupervised machine learning technique, namely K-medoids clustering, to compress the set of nighttime propagation directions in angular space. Even after clustering, the resulting ensemble of representative paths remains large. Consequently, we implement the MSW resonance propagation scheme based on fast Strang splitting~\cite{Strang:1968}, which decomposes the Hamiltonian into vacuum and matter components in the mass basis and reformulates the matter term using projector algebra to enable efficient parallel computation. This operator-splitting strategy significantly reduces computational complexity, facilitating rapid scanning over $10^4$–$10^5$ propagation trajectories and a dense energy grid of 2000 points spanning $0.1$–$16~\mathrm{MeV}$, while maintaining numerical accuracy in oscillation probabilities at the $\mathcal{O}(10^{-4})$ level or better.
 
In the numerical applications presented below, the $^{8}\mathrm{B}$ component is adopted as a representative solar-neutrino case, since the observable modulation of solar-neutrino oscillations induced by matter effects is most pronounced in this comparatively high-energy component.
At the conclusion of this paper, we present site-specific flux predictions, which are not only for the China Jinping Underground Laboratory (CJPL) but also for other underground laboratories currently conducting or planning solar neutrino research programs—incorporating corrections for both diurnal asymmetry and the regeneration probability distribution arising from matter effects.

\section{Sun–Earth geometry and source-path construction}
\subsection{Approximate Analytical Approach: the time-dependent detector-to-Sun vector}
\label{sec:meeus}
As a low-order analytical cross-check of the ephemeris-based geometry, we construct an approximate geometric detector-to-Sun vector using a Keplerian/Meeus-style solar model.  
We deliberately omit apparent astronomical effects, such as light-time corrections and relativistic aberration, to preserve simplicity and transparency in the geometric formulation. In this subsection, we derive the detector-to-Sun unit vector \(\hat{\mathbf{s}}_{\rm ENU}(t)\), expressed in the local east–north–up (ENU) frame, together with the instantaneous distance \(\mathcal{R}_{\odot d}(t)\) between the detector and the Sun's center.  

Within the Meeus solar-coordinate algorithm~\cite{Meeus1998AstronomicalAlgorithms}, the geometric solar longitude \(\lambda_\odot(t)\), the Earth–Sun distance \(R_{\oplus\odot}(t)\), and the mean obliquity of the ecliptic \(\epsilon(t)\) are computed via low-order polynomial expansions in terms of the Julian century:  
\begin{equation}
T(t) = \frac{JD(t) - 2451545.0}{36525},
\end{equation}  
where \(JD(t)\) denotes the Julian Day number corresponding to the observation time \(t\). Throughout this subsection, we use
\(\lambda_\odot(t)\), \(R_{\oplus\odot}(t)\), and \(\epsilon(t)\)
as shorthand for
\(\lambda_\odot[T(t)]\), \(R_{\oplus\odot}[T(t)]\), and
\(\epsilon[T(t)]\), respectively.
Neglecting the solar ecliptic latitude in this approximation, the corresponding equatorial coordinates are  
\begin{align}
\alpha_\odot(t)
&=
\operatorname{atan2}
\big[
\cos\epsilon(t) \sin\lambda_\odot(t),\,
\cos\lambda_\odot(t)
\big],\\
\delta_\odot(t)
&=
\arcsin
\big[
\sin\epsilon(t) \sin\lambda_\odot(t)
\big].
\end{align}  
For a detector located at geodetic latitude \(\phi\), east-positive longitude \(\ell\), and altitude \(h\), the local hour angle is given by  
\begin{equation}
H(t)=\theta_{\rm GMST}(t)+\ell-\alpha_\odot(t),
\end{equation}  
where \(\theta_{\rm GMST}\) denotes the Greenwich Mean Sidereal Time angle. The unit vector pointing from the Earth’s center to the Sun’s center, expressed in the local east–north–up (ENU) frame, is given by  
\begin{equation}
\hat{\mathbf{n}}_{\oplus\odot,\,\mathrm{ENU}}(t) =
\begin{pmatrix}
-\cos\delta_\odot(t)\sin H(t) \\
\cos\phi\,\sin\delta_\odot(t)
-\sin\phi\,\cos\delta_\odot(t)\cos H(t) \\
\sin\phi\,\sin\delta_\odot(t)
+\cos\phi\,\cos\delta_\odot(t)\cos H(t)
\end{pmatrix}.
\end{equation}
Defining the effective detector radius as \(R_d = R_\oplus + h\), the vector from the detector to the Sun’s center may be written as  
\begin{align}  
&\mathbf{r}_{\odot d}(t) = R_{\oplus\odot}(t)\,\hat{\mathbf{n}}_{\oplus\odot,\,\mathrm{ENU}}(t) - R_d  
\begin{pmatrix}  
0 \\ 0 \\ 1  
\end{pmatrix}, \nonumber \\  
&\hat{\mathbf{s}}_{\mathrm{ENU}}(t) = \frac{\mathbf{r}_{\odot d}(t)}{\|\mathbf{r}_{\odot d}(t)\|}, \qquad  
\mathcal{R}_{\odot d}(t) = \|\mathbf{r}_{\odot d}(t)\|,  
\end{align}  
where \(\mathcal{R}_{\odot d}(t)\) denotes the instantaneous distance between the detector and the Sun's center, and \(\hat{\mathbf{s}}_{\mathrm{ENU}}(t)\) is the corresponding unit vector, expressed in the local ENU frame—pointing from the detector toward the Sun.

\subsection{High-precision geometric ephemeris: the time-dependent detector-to-Sun vector}
\label{sec:astropy}
We construct the Sun–Earth geometry using the barycentric Cartesian positions of the Sun and Earth, as computed by Astropy~\cite{Astropy}. Specifically, the geocentric Earth-to-Sun vector is obtained by subtracting the Earth's barycentric position from that of the Sun, both expressed in the International Celestial Reference System (ICRS) relative to the Solar System barycenter. This approach intentionally omits implicit, observer-dependent apparent-position corrections, such as light-time delay and relativistic aberration, thereby ensuring full explicit control over such effects. Should source-point-dependent corrections become necessary in a future three-dimensional solar-source modeling framework, they can be consistently applied at the source-geometry level, thereby avoiding double-counting and eliminating potential conflicts with hidden apparent-position conventions.

The detector's position is specified by its geodetic longitude, latitude, and altitude, initially defined in the International Terrestrial Reference System (ITRS). For each time sample, we compute the celestial-to-terrestrial rotation matrix using Astropy's time-handling infrastructure, IERS Earth-orientation parameters, and the ERFA routine \texttt{c2t06a}. The transpose of this matrix transforms the detector's geocentric position vector from the ITRS into the GCRS. The geocentric Earth-to-Sun vector obtained by subtracting the barycentric ephemeris positions is expressed in an ICRS/BCRS-aligned celestial Cartesian basis. Since the GCRS uses the same spatial orientation, the two geocentric vectors can be consistently combined at the geometric precision adopted here. The resulting Sun--detector vector is then rotated back into the ITRS using the original celestial-to-terrestrial rotation matrix.

As the time-varying detector-to-Sun geometries are determined, we compute the distance between the detector and the solar center, denoted $\mathcal{R}_{\odot d}(t)$, as well as the unit vector pointing from the detector to the solar center, $\hat{\mathbf{s}}_{\rm ENU}(t)$, expressed in the local east–north–up (ENU) reference frame. From this local direction vector, we define the solar azimuth angle $az(t)$, the nadir angle $\eta(t)$, and a Boolean nighttime flag $\texttt{is\_night}(t)$. Specifically, nighttime samples correspond to instances when the Sun lies below the local horizon; for such samples, $\eta(t) \in [0^\circ, 90^\circ]$. The resulting time-series  
\begin{equation}
\big[\mathcal{R}_{\odot d}(t_i),\ \eta(t_i),\ az(t_i),\ \texttt{is\_night}(t_i)\big],
\end{equation}
serves as input to the Earth-crossing trajectory construction and representative-path clustering procedures described in subsequent sections.

As an independent validation, we reconstruct the same geometric quantities using Skyfield~\cite{rhodes2019skyfield}. In this implementation, the geocentric Earth-to-Sun vector is derived from the instantaneous vector difference between the Sun's and Earth's ephemeris positions—rather than relying on Skyfield's observing pipeline methods \texttt{observe()} or \texttt{apparent()}. The terrestrial-frame transformation is then performed using Skyfield's International Terrestrial Reference System (ITRS) rotation model. The maximum absolute discrepancies between the two high-precision implementations are on the order of tens of meters for both the derived detector-to-Sun distance and the Earth-crossing chord length. Additionally, we compare a low-order analytical model—based on Keplerian and Meeus' approximations—against the Astropy-based geometric backend: the discrepancy in spherical-Earth chord length remains below $2\,{\rm km}$, while the detector-to-Sun distance differs by less than $\sim \mathcal{O}(10^4)\,{\rm km}$.

\subsection{Point-cloud representation of the finite solar neutrino production region}

After obtaining the time-dependent detector-to-Sun vector, we further incorporate the finite spatial extent of the solar neutrino production region. Rather than modeling the solar source as a point-like emitter located precisely at the solar center, we represent it as a compact angular point cloud centered on the direction from the detector to the solar center. This point-cloud representation yields the finite-source correction subsequently employed in path-space clustering.

Let $\rho_j$ denote the projected radial coordinate of the $j$-th source node, expressed in units of the solar radius $R_\odot$. The projected source nodes are drawn from the cumulative distribution function of the projected solar neutrino production profile, evaluated at a reference Sun–detector distance. For each time sample $t_i$, the angular radius corresponding to the $j$-th projected node is given by  
\begin{equation}
q_j(t_i)
=
\arctan\left(
\frac{\rho_j R_\odot}{\mathcal{R}_{\odot d}(t_i)}
\right),
\label{eq:q_nodes_finite_source}
\end{equation}
where $\mathcal{R}_{\odot d}(t_i)$ denotes the instantaneous distance between the detector and the solar center.

The projected source weight associated with each annulus is then computed via numerical line-of-sight projection of the tabulated, spherically symmetric solar neutrino production profile. Specifically, the production density is sampled along the line of sight, interpolated from the underlying solar model table, and integrated over the corresponding projected annular region. We denote the resulting normalized weight by $w_j(t_i)$, satisfying  
\begin{equation}
\sum_j w_j(t_i)=1.
\label{eq:source_weight_normalization}
\end{equation}

Since both the solar density model and the solar neutrino production model are radially symmetric, the resulting finite-source angular distribution is isotropic about the central detector–Sun direction. From the perspective of the detector, the finite solar production region subtends a small angular extent on the unit sphere of incoming neutrino directions.  

In the local tangent plane, we therefore characterize its transverse angular width via the second moment  
\begin{equation}  
\sigma_{\rm tan}^2(t_i)  
=  
\frac{1}{2}\sum_j w_j(t_i)\, q_j^2(t_i),  
\label{eq:sigma_tan_definition}  
\end{equation}  
where the factor $1/2$ follows from transverse isotropy, since $\langle q^2\rangle=\langle x^2+y^2\rangle =2\sigma_{\tan}^2$.

\begin{figure*}[!htbp]
    \centering
    \includegraphics[width=0.98\textwidth]{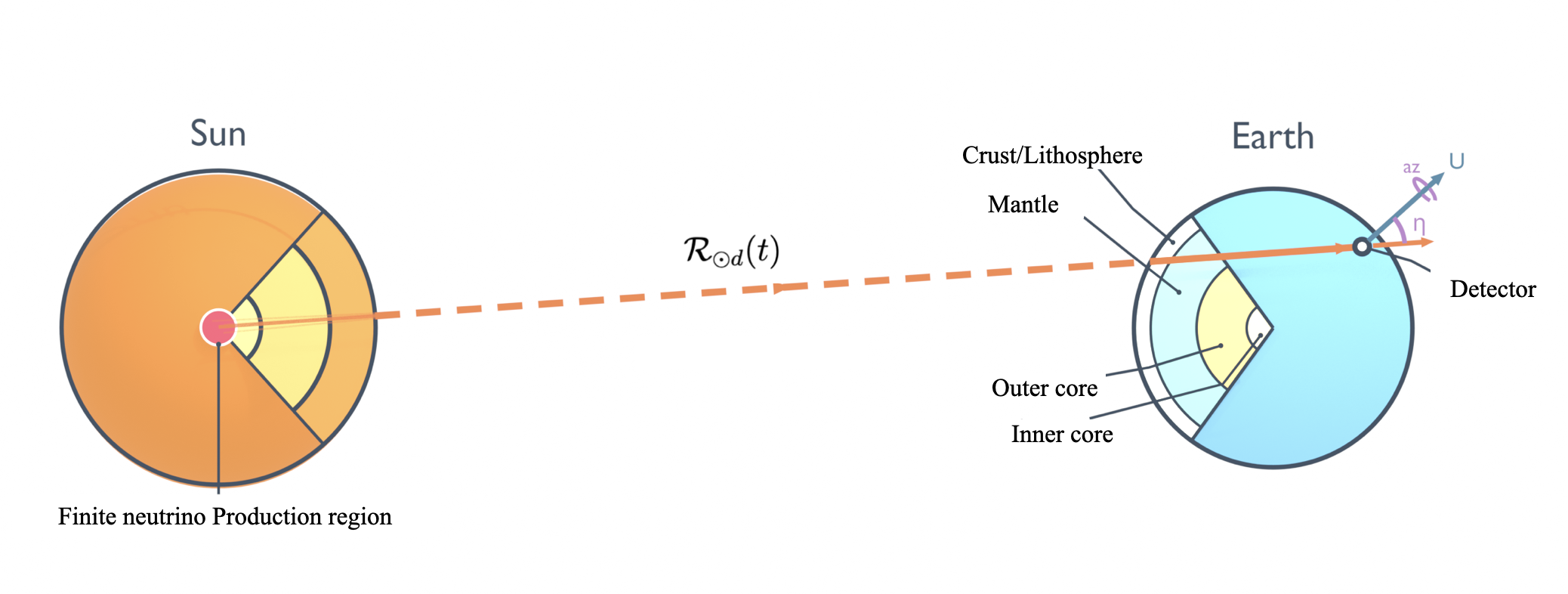}
    \caption{
    Schematic illustration of the solar neutrino source and propagation geometry. The finite neutrino production region within the Sun generates a narrow ensemble of neutrino source trajectories. The dashed segment represents vacuum propagation over the time-dependent Sun–detector distance $\mathcal{R}_{\odot d}(t)$, whereas the solid segment depicts a representative nighttime trajectory traversing the Earth to reach the underground detector. The terrestrial cross-section highlights the crust/lithosphere, mantle, outer core, and inner core—regions incorporated into the hybrid three-dimensional electron-density model. At the detector location, $\hat{\mathbf{U}}$ denotes the local upward direction. The nadir angle $\eta$ is defined equivalently as the angle between $\hat{\mathbf{U}}$ and the neutrino propagation direction; $az$ denotes the azimuthal angle measured about the local vertical axis. The relative sizes and separation of the Sun and Earth are schematic and not drawn to scale.
    }
    \label{fig:sun_earth_geometry}
\end{figure*}

Figure~\ref{fig:sun_earth_geometry} summarizes the source and propagation geometry used in this work, including the finite neutrino production region, the time-dependent Sun--detector baseline, a representative nighttime Earth-crossing trajectory, and the local angular convention at the detector. The geometrical record employed in the representative-path construction is thus  
\begin{equation}  
\left[  
\mathcal{R}_{\odot d}(t_i),\  
\eta(t_i),\  
az(t_i),\  
\texttt{is\_night}(t_i),\  
\sigma_{\rm tan}(t_i)  
\right].  
\label{eq:geometry_record_with_sigma}  
\end{equation}
\section{Earth-side MSW propagation}
\subsection{Terrestrial electron-density model}
The terrestrial electron-density model employed in this work follows the same construction as that adopted in our previous reactor-neutrino study~\cite{2026ReactorEarth}. We therefore summarize only those components directly relevant to the present solar-neutrino calculation. The charged-current matter potential is given by  
\begin{equation}  
V_e(\mathbf{x}) = \sqrt{2}\,G_F\,N_e(\mathbf{x}),  
\end{equation}  
where the electron number density $N_e(\mathbf{x})$ is derived using the identical density and electron-fraction prescriptions as in Ref.~\cite{2026ReactorEarth}.  

For the one-dimensional reference calculation, we adopt the Preliminary Reference Earth Model (PREM) radial density profile. For the three-dimensional calculation, we employ a hybrid terrestrial model: LITHO1.0~\cite{LITHO10} is used in the shallow crustal region, SAW642AN~\cite{SAW642AN} in the mantle, and PREM~\cite{PREM} in the deep-Earth region (including the outer and inner core). The resulting electron-density field is interpolated along each Earth-crossing trajectory and converted into a sequence of midpoint matter potentials, which serve as the Earth-model input to the subsequent neutrino propagation solver. No new geophysical modeling assumptions are introduced in this work; the novelty lies solely in the application of this established terrestrial model to time-dependent solar-neutrino nighttime trajectories and their associated representative-path compression.

\subsection{Evolution equation}
The MSW effect governing neutrino propagation is described by a Schr\"odinger-like equation along the trajectory through matter. In the vacuum mass basis, the evolution is given by  
\begin{equation} 
i\,\frac{\text{d}}{\text{d}x}\,|\psi(x)\rangle \;=\; H(x;E)\,|\psi(x)\rangle, 
\label{eq:schrodinger_evolution}
\end{equation}  
where $E$ denotes the neutrino energy and $|\psi(x)\rangle$ is the quantum state vector.  

It is convenient to characterize the propagation using the evolution operator $S(x,x_0)$, defined via  
\begin{equation} 
|\psi(x)\rangle \;=\; S(x,x_0)\,|\psi(x_0)\rangle, 
\qquad 
i\,\frac{\text{d}}{\text{d}x}\,S(x,x_0) \;=\; H(x;E)\,S(x,x_0), 
\qquad 
S(x_0,x_0)=I. 
\label{eq:evolution_operator_def}
\end{equation}  
Since $H(x; E)$ generally fails to commute at different positions $x$, the formal solution takes the form of a path-ordered exponential:  
\begin{equation} 
S(L,0) \;=\; \mathcal{P}\exp\!\left[-i\int_{0}^{L} H(x;E)\,\text{d}x\right], 
\label{eq:path_ordered_exp}
\end{equation}  
where $\mathcal{P}$ denotes path ordering with respect to the coordinate $x$, and $L$ represents the total propagation length—e.g., through the Earth or the Sun.  

In the vacuum mass basis, the Hamiltonian decomposes as  
\begin{equation}
H(x;E)
=
H_{\rm vac}(E)+H_{\rm mat}(x),
\label{eq:H_decomp}
\end{equation}  
with  
\begin{equation}
H_{\rm vac}(E)
=
\mathrm{diag}\!\left(
0,\,
\frac{\Delta m_{21}^2}{2E},\,
\frac{\Delta m_{31}^2}{2E}
\right),
\qquad
H_{\rm mat}(x)
=
U^\dagger
\mathrm{diag}\!\left(V_e(x),0,0\right)
U.
\label{eq:Hvac_Hmat_mass}
\end{equation}  
Here, $U$ is the leptonic mixing matrix (the PMNS matrix), and $V_e(x)=\sqrt{2}\, G_F\, N_e(x)$ is the charged-current matter potential, where $G_F$ is the Fermi coupling constant and $N_e(x)$ is the electron number density at position $x$.

For the Earth-side calculation, we approximate the Earth's matter density profile using \(N\) piecewise-constant layers. Within the \(n\)-th layer, the matter potential is evaluated at the midpoint \(x_n\) of that layer, and the Hamiltonian is approximated as  
\begin{equation}  
H_n(E) = H_{\rm vac}(E) + H_{\rm mat}(x_n).  
\end{equation}  
Under this layer-wise discretization of the Earth’s matter profile, the path-ordered exponential in Eq.~\eqref{eq:path_ordered_exp} is approximated by the ordered product  
\begin{align}  
S^\oplus(E;\ x_N=L_\oplus,0)  
&\approx \prod_{n=N-1}^{0} \exp\!\left[-iH_n(E)\Delta x_n\right]  
\nonumber\\  
&= e^{-iH_{N-1}(E)\Delta x_{N-1}} \cdots e^{-iH_1(E)\Delta x_1} e^{-iH_0(E)\Delta x_0}.  
\label{eq:earth_layer_product}  
\end{align}  
This piecewise-constant exponential product serves as the foundation for the Earth-side Strang-splitting implementation described below. In contrast, the solar-side calculation employs the same formal evolution equation Eq.~\eqref{eq:path_ordered_exp}, but utilizes a distinct high-order local discretization scheme based on the commutator-free fourth-order Magnus expansion.

\subsection{The Strang-splitting Approach and Closed-form Exponential of a Rank-one Matrix}
To evaluate Eq.~\eqref{eq:earth_layer_product}, a straightforward approach is to compute the matrix exponential  
$\exp[-iH_n(E)\Delta x_n]$ in each layer $n$, for example, via spectral decomposition of the Hermitian matrix $H_n(E)$. However, this method becomes computationally expensive when applied to large ensembles of trajectories and dense energy grids. We therefore adopt the symmetric, second-order Strang splitting~\cite{Strang:1968},  
\begin{align}
S^\oplus(E;\Delta x_n)&=\exp\!\big[-i\big(H_{\rm vac}(E)+H_{{\rm mat},n}\big)\Delta x_n\big]
\; \nonumber \\
&\approx\;
e^{-iH_{\rm vac}\Delta x_n/2}\,
e^{-iH_{{\rm mat},n}\Delta x_n}\,
e^{-iH_{\rm vac}\Delta x_n/2},
\label{eq:strang_layer}
\end{align}  
which achieves global second-order accuracy (with local truncation error $\mathcal{O}(\Delta x_n^3)$) while preserving unitarity up to machine precision.  

Using Eq.~\eqref{eq:Hvac_Hmat_mass}, the exponential of $H_{{\rm mat},n}$ admits a closed-form expression:  
\begin{align}
e^{-iH_{{\rm mat},n}\Delta x_n}
&= e^{-iU^\dagger \mathrm{diag}\!\left(V_n,0,0\right)U\,\Delta x_n} \nonumber\\
&= e^{-iV_n\Delta x_n\,U^\dagger |\nu_e\rangle_f {\ }_f\langle \nu_e|\,U} \nonumber\\
&= e^{-iV_n\Delta x_n\,|e\rangle\langle e|} \nonumber\\
&= I+\left(e^{-iV_n\Delta x_n}-1\right)|e\rangle\langle e|,
\label{eq:rank1_projector}
\end{align}  
where we define the electron-flavor state in the mass basis as  
$|e\rangle \equiv U^\dagger|\nu_e\rangle_{f} = (U_{e1}^*,U_{e2}^*,U_{e3}^*)^T$,  
and exploit the projector identity $P^m=P$ for all integers $m\ge1$, with $P\equiv |e\rangle\langle e|$.

By combining Eqs.~\eqref{eq:strang_layer} and \eqref{eq:rank1_projector}, the layer update in the mass basis admits an explicit implementation.  
The vacuum half-step is diagonal,  
\begin{equation}  
e^{-iH_{\rm vac}\Delta x_n/2}  
=  
\mathrm{diag}\!\left(1,\ e^{-i\phi_{21,n}},\ e^{-i\phi_{31,n}}\right),  
\qquad  
\phi_{ij,n}\equiv \frac{\Delta m_{ij}^2\,\Delta x_n}{4E}.  
\label{eq:vac_halfstep_phase}  
\end{equation}  
The matter step takes the form of a rank-one update,  
\begin{equation}  
|\psi\rangle \;\leftarrow\; |\psi\rangle  
+\left(e^{-iV_n\Delta x_n}-1\right)\,|e\rangle\,\langle e|\psi\rangle,  
\label{eq:matter_rank1_update}  
\end{equation}  
so that a complete Strang-splitting layer update reads  
\begin{equation}  
|\psi_{n+1}\rangle \;=\;  
e^{-iH_{\rm vac}\Delta x_n/2}\,  
\Big[I+\left(e^{-iV_n\Delta x_n}-1\right)|e\rangle\langle e|\Big]\,  
e^{-iH_{\rm vac}\Delta x_n/2}\,  
|\psi_n\rangle.  
\label{eq:strang_layer_explicit}  
\end{equation}  

In practice, each layer update comprises two diagonal phase multiplications and one rank-one update.  
Defining $a_n \equiv \langle e|\psi_n\rangle$, the matter step simplifies to  
$\psi \leftarrow \psi + (e^{-iV_n\Delta x_n}-1)\,a_n\,|e\rangle$.  
Consequently, the computational cost scales as $\mathcal{O}(N_{\rm layers}\, N_E)$ with a small prefactor, and the scheme naturally accommodates non-uniform step sizes $\Delta x_n$.

\section{Solar-side MSW propagation} 
\label{sec:solar_msw}
\subsection{Solar model inputs and the neutrino source distribution}
The solar-side calculation depends on three external solar inputs: the solar matter density profile, the spatial distribution of neutrino production, and the undistorted neutrino energy spectrum. In this study, we adopt the Bahcall BS05(OP)~\cite{Bahcall:2004kjs} solar model as our reference. The radial solar structure is extracted from the file \texttt{bs05op.dat}, while the neutrino production profile is obtained from \texttt{bs2005opflux.dat}. The normalized $^{8}\mathrm{B}$ neutrino energy spectrum is taken from \texttt{b8spectrum.txt}~\cite{Bahcall:1996qv}~\cite{BahcallSolarData}. We select the $^{8}\mathrm{B}$ component as the representative case, as the observable modulation of solar-neutrino oscillations induced by matter effects is most prominent in this relatively high-energy component.

Collectively, these inputs determine: (i) the electron matter potential along each solar trajectory; (ii) the production-weighting function used in the source integral; and (iii) the energy-dependent weighting applied in the final flux prediction.  

Critically, the subsequent propagation algorithm is not specific to the $^{8}\mathrm{B}$ neutrinos and remains fully decoupled from the particular solar input tables. By substituting the corresponding production profiles and energy spectra, the time-varying propagation geometry, finite-source averaging, and oscillation effects for other solar-neutrino components can be evaluated straightforwardly within the same computational framework. Likewise, alternative solar models or updated spectral inputs can be incorporated without any modification to the core propagation algorithm.

\subsection{Backward propagation}
\label{sec:back_prop}
The adopted solar density model and the neutrino production profile are both spherically symmetric. Along a fixed impact-parameter chord, distinct neutrino production points correspond to different initial positions within the \(^{8}\mathrm{B}\) source region. If propagation is carried out in the forward direction—i.e., from each production point outward to the solar surface—each point necessitates an independent evolution, resulting in numerous overlapping path segments across different trajectories. Such overlap renders many of these forward calculations computationally redundant.  

To eliminate this redundancy, we perform the entire propagation in reverse. For each impact parameter \(b\), we initiate the calculation at the solar surface with the identity matrix and propagate inward along the chord. As the trajectory enters the \(^{8}\mathrm{B}\) source region, the evolution operator is projected onto the electron-neutrino flavor basis and weighted by the local \(^{8}\mathrm{B}\) production density. Consequently, all production points sharing the same impact parameter benefit from a single backward propagation, thereby substantially reducing the computational cost associated with the solar-side source integral.

\subsection{Commutator-free fourth-order Magnus propagation}
The solar-side calculation begins from the same path-ordered evolution operator given in Eq.~\eqref{eq:path_ordered_exp}. However, due to the significantly longer solar baseline and the consequent need for stringent control over accumulated discretization error, the piecewise-constant layer product employed in Eq.~\eqref{eq:earth_layer_product} is not used. Instead, each local segment is evaluated using a commutator-free fourth-order Magnus integrator~\cite{Magnus}.

In this solar-side treatment, we denote by \(E_k\) the \(k\)-th point on the discretized energy grid. This explicit energy indexing proves advantageous, as the solar CF4 coefficients, the adaptive-mesh predictor, and the precomputed Ohlsson–Snellman quantities all depend jointly on both energy and layer indices. In contrast, the preceding Earth-side discussion was formulated for a generic energy \(E\), rendering additional energy-index notation unnecessary.

A standard commutator-free fourth-order Magnus approximation~\cite{BlanesMoan2006} to the evolution operator takes the form  
\begin{equation}
S^\odot(E_k, \Delta x_n)=\exp{\big[-i\Delta x_n(a_1H_{1,n,k}+a_2H_{2,n,k})\big]}\exp{\big[-i\Delta x_n(a_2H_{1,n,k}+a_1H_{2,n,k})\big]},
\label{eq:CF4}
\end{equation}
where  
\begin{equation}
H_{i,n,k}=H_{vac}(E_k)+H_{mat}\big(x_n+c_i\Delta x_n\big), \quad i=1,2. 
\end{equation}  
The parameters are given by \(a_{1,2} = \frac{1}{4} \pm \frac{\sqrt{3}}{6}\) and \(c_{1,2} = \frac{1}{2} \mp \frac{\sqrt{3}}{6}\). Here, \(H_{mat}\big(x_n+c_i\Delta x_n\big)\) denotes the Gaussian quadrature nodes used to sample the matter potential within each discretized layer step. Specifically, \(H_{mat}\big(x_n+c_i\Delta x_n\big) = U^\dagger\text{diag}\big(V^{CC}_{i,n},\, 0,\, 0\big)U\), where \(V^{CC}_{i,n} = V^{CC}\big(x_n+c_i\Delta x_n\big)\). By leveraging the rank-one projector property of $H_{\text{mat}}$, we have  
\begin{align}
H_{\text{mat},i,n} &= U^\dagger\ \text{diag}\big(V^{CC}_{i,n},\, 0,\, 0\big)\, U \nonumber \\
&= V^{CC}_{i,n}\, U^\dagger 
\begin{pmatrix}
1 \\ 0 \\ 0
\end{pmatrix}
\begin{pmatrix}
1 & 0 & 0
\end{pmatrix}
U \nonumber \\
&= V^{CC}_{i,n}\, |e\rangle\langle e| \nonumber \\
&= V^{CC}_{i,n}\, P,
\end{align}
where $U$ denotes the PMNS mixing matrix, and $|e\rangle = U^\dagger |\nu_e\rangle_{\text{flavor}}$ represents the electron-flavor state projected onto the vacuum mass basis—equivalently, the projection operator onto the electron-flavor direction in that basis. Using this identity, Eq.~\eqref{eq:CF4} can be recast as  
\begin{align}
S^\odot(E_k, \Delta x_n)=\exp{\bigg[-i\Delta x_n\bigg(\frac{1}{2} H_{\text{vac}} + V_{\text{eff},1,n}\,P\bigg)\bigg]}\,
\exp{\bigg[-i\Delta x_n\bigg(\frac{1}{2} H_{\text{vac}} + V_{\text{eff},2,n}\,P\bigg)\bigg]},
\label{eq:CF4_2stage}
\end{align}
with $V_{\text{eff},1,n}=a_1 V^{CC}_{1,n}+a_2 V^{CC}_{2,n}$ and $V_{\text{eff},2,n}=a_2 V^{CC}_{1,n}+a_1 V^{CC}_{2,n}$. We refer to Eq.~\eqref{eq:CF4_2stage} as the *CF4 two-stage form*. Although this form can be formally combined through a truncated Baker--Campbell--Hausdorff expansion, this introduces an additional approximation beyond the original CF4 discretization. We therefore retain the two-stage form throughout this work.

\subsection{Ohlsson-Snellman roots and Cayley-Hamilton exponentiation}
A direct evaluation of the matrix exponential at every propagation step, which uses a generic \texttt{expm} routine or an eigensolver, is computationally inefficient for the vast number of energy–path–step combinations required in solar neutrino calculations. Such general-purpose routines are designed for arbitrary matrices and typically rely on iterative algorithms or relatively expensive linear-algebra operations, making them poorly suited for massively parallel GPU execution.

An alternative approach is to approximate the exponential associated with each CF4 stage via matrix-splitting formulas—such as the Strang or Yoshida splittings. However, this strategy is fundamentally incompatible with the Magnus-based CF4 method: standard splitting schemes require smaller step sizes to control the additional Baker–Campbell–Hausdorff (BCH) error introduced within each exponential evaluation, whereas the Magnus CF4 method is specifically designed to permit relatively large step sizes over the extremely long solar baseline while maintaining a prescribed accuracy.

A further practical concern arises from numerical experimentation. Although Yoshida splitting is formally fourth-order and thus appears asymptotically compatible with the CF4 scheme, our tests reveal that it can induce step-size resonance in the low-energy regime of solar neutrino propagation. Once triggered, this resonance leads to significant error amplification that the fifth-order local splitting error cannot adequately suppress.

Instead, we exploit the special rank-one structure of the matter Hamiltonian in the mass basis. Specifically, we adopt the Ohlsson–Snellman formalism~\cite{Ohlsson:1999um}, which decomposes the Hamiltonian into its trace and traceless parts:
\begin{align}
H=\frac{1}{3}\,\mathrm{Tr}[H]\,I + T,
\end{align}
where $I$ denotes the identity matrix. Since the global phase factor does not influence MSW propagation, the final expression simplifies to  
\begin{align}
\widetilde S^\odot(E_k,\Delta x_n)&=
\exp{\bigg[-i\Delta x_n\big( \boldsymbol{\Delta}_k+V_{\text{eff},1,n}Q\big)\bigg]}\exp{\bigg[-i\Delta x_n\big( \boldsymbol{\Delta}_k+V_{\text{eff},2,n}Q\big)\bigg]} \nonumber \\
& =\exp{\big(-i\Delta x_nT_{1,k,n}\big)}\exp{\big(-i\Delta x_nT_{2,k,n}\big)}.
\label{eq:CF4_2stage_OS}
\end{align}
Here, $T_{i,k,n}\equiv \boldsymbol{\Delta}_k+V_{\text{eff},i,n}Q$, with $i=1,2$, and 
\begin{align}
\boldsymbol{\Delta}_k=H_{vac}(E_k)-\frac13\mathrm{Tr}\big[H_{vac}(E_k)\big]I,\quad
Q=P-\frac1 3 I.
\end{align}

In the LMA region of solar-neutrino oscillation parameters, no near-degenerate eigenvalue configurations arise; consequently, evaluation via the closed-form solution of the associated cubic equations remains numerically stable and is particularly well suited for GPU-parallelized computation. Moreover, since the matrices $T_{i,k,n}$ in Eq.~\eqref{eq:CF4_2stage_OS} are traceless, the characteristic cubic equation for their eigenvalues—evaluated at fixed neutrino energy $E_k$ and layer index $n$—takes the form  
\begin{align}
\lambda^3 - p_{i}\lambda-q_{i}=0,
\label{eq:cubic}
\end{align}
whose analytical solution is given by  
\begin{align}
&\lambda_{m,i} = \rho_i \cos\!\left(\theta_i - \frac{2m\pi}{3}\right),\quad m=0,1,2,  \\
&\rho_i = 2\sqrt{\frac {p_{i}}{3}},\quad \theta_i =\frac{1}{3} \arccos(\xi_i),\quad \xi_i = \frac{3\sqrt{3}\,q_{i}}{2\,p_{i}^{3/2}}.\nonumber
\end{align}
The coefficients $p_i$ and $q_i$ in Eq.~\eqref{eq:cubic} are expressed in terms of matrix traces as  
\begin{align}
&p_{i} = \frac{1}{2} \operatorname{Tr}\!\left[T_{i}^2\right], \nonumber \\
&q_{i} = \frac{1}{3} \operatorname{Tr}\!\left[T_{i}^3\right].
\label{eq:cubic_parameters}
\end{align}
To further optimize computational efficiency, Eq.~\eqref{eq:cubic_parameters} can be decomposed into energy-dependent and potential-dependent contributions. Specifically, the energy-dependent terms—namely $A$, $B$, $C$, and $D$ defined below—can be precomputed before the propagation step:  
\begin{align}
&A = \frac{1}{2} \operatorname{Tr}\!\left[\boldsymbol{\Delta}_k^2\right], \nonumber \\
&B = \operatorname{Tr}\!\left[\boldsymbol{\Delta}_k \cdot P\right], \nonumber \\
&C =\det\!\left[\boldsymbol{\Delta}_k\right], \nonumber \\
&D = \operatorname{Tr}\!\left[\boldsymbol{\Delta}_k^2\cdot P\right]-\frac{1}{3}\operatorname{Tr}\!\left[\boldsymbol{\Delta}_k^2\right].
\label{eq:cubic_abcd}
\end{align}
In this case, Eq.~\eqref{eq:cubic_parameters} simplifies to  
\begin{align}
&p_{i} = A + B V_{\text{eff},i} + \frac{1}{3} V_{\text{eff},i}^2, \nonumber \\
&q_{i} = C + D V_{\text{eff},i} + \frac{1}{3} B V_{\text{eff},i}^2 + \frac{2}{27} V_{\text{eff},i}^3.
\label{eq:cubic_parameters_abcd}
\end{align}  
Upon computing the eigenvalues, we employ the Cayley–Hamilton theorem to decompose Eq.~\eqref{eq:CF4_2stage_OS} as  
\begin{align}
\widetilde S^\odot &= \exp\big(-i\Delta x_n T_{1}\big)\,\exp\big(-i\Delta x_n T_{2}\big) \nonumber \\
&= \big(\alpha_{0}^{(1)}I + \alpha_{1}^{(1)}T_{1} + \alpha_{2}^{(1)}T^2_{1}\big)
\big(\alpha_{0}^{(2)}I + \alpha_{1}^{(2)}T_{2} + \alpha_{2}^{(2)}T^2_{2}\big),
\label{eq:CF4_2stage_CH}
\end{align}  
where the Cayley–Hamilton coefficients are given by  
\begin{align}
\alpha_{0}^{(i)}
&=
\sum_{m=0}^{2}
\frac{
\big(\lambda_{m,i}^2 - p_i\big)
e^{-i\Delta x_n\lambda_{m,i}}
}
{3\lambda_{m,i}^2 - p_i},
\nonumber\\
\alpha_{1}^{(i)}
&=
\sum_{m=0}^{2}
\frac{
\lambda_{m,i}\,
e^{-i\Delta x_n\lambda_{m,i}}
}
{3\lambda_{m,i}^2 - p_i},
\nonumber\\
\alpha_{2}^{(i)}
&=
\sum_{m=0}^{2}
\frac{
e^{-i\Delta x_n\lambda_{m,i}}
}
{3\lambda_{m,i}^2 - p_i}.
\label{eq:CF4_2stage_CH_coef}
\end{align}  
All quantities appearing in the coefficients $\alpha_\ell^{(i)}$ are obtained from closed-form algebraic expressions. Consequently, evaluating each CF4 stage exponential involves only a fixed sequence of scalar arithmetic operations and low-dimensional matrix multiplications—no iterative diagonalization, convergence monitoring, or adaptive sub-stepping is required. Moreover, the energy-dependent parameters $A$, $B$, $C$, and $D$ can be precomputed before time propagation, while layer dependence enters exclusively through the effective potentials $V_{\text{eff}, i}$. This fixed computational cost and regular data access pattern render the analytic Cayley–Hamilton evaluator especially well suited for massively parallel GPU implementation. Compared with a generic \texttt{expm} routine or a splitting-based approximation, this construction directly evaluates the CF4-stage exponentials and thereby avoids introducing additional inner-splitting errors.

\subsection{Adaptive mesh construction with a two-flavor predictor}
The guiding principle of this section is that the adaptive step-size method is fundamentally an empirical algorithm, wherein the desired target accuracy primarily dictates the choice of tolerance. Consequently, it is unnecessary—and computationally inefficient—to employ full $3 \times 3$ matrix diagonalization for determining the adaptive mesh. For the solar-neutrino problem under consideration, the dominant matter-induced evolution arises predominantly from the (1–2) flavor mixing mechanism. We therefore adopt a computationally economical two-flavor predictor, in which the (1–2) active subblock is exponentiated analytically using the closed-form $SU(2)$ solution.

\subsubsection{The tolerance criterion}  
The tolerance is imposed on an operator-level error indicator—rather than on the state-level error resulting from application of the propagator to a specific incoming state. This distinction is crucial: the incoming state at any given mesh cell already encodes the cumulative propagation history across all preceding cells. To ensure that the mesh construction remains independent of this history, we estimate the worst-case component-wise error induced by the local defect operator.  

Let $\Delta \widetilde{S}$ denote the local defect operator obtained via step-doubling comparison. For an arbitrary incoming vector $v = (a,b,c)^T$, the $i$-th component of the defect satisfies  
\begin{equation}  
\left|(\Delta \widetilde{S}v)_i\right|  
=  
\left|\Delta \widetilde{S}_{i1}\,a + \Delta \widetilde{S}_{i2}\,b + \Delta \widetilde{S}_{i3}\,c\right|  
\leq  
\left(\sum_{j=1}^{3} |\Delta \widetilde{S}_{ij}|^2\right)^{1/2}  
\left(\sum_{j=1}^{3} |v_j|^2\right)^{1/2}.  
\end{equation}  
For a normalized incoming state ($\|v\|_2 = 1$), the maximum possible error in the $i$-th component is thus bounded by the Euclidean norm of the $i$-th row of $\Delta \widetilde{S}$. Accordingly, we define the local error indicator as  
\begin{equation}  
\|\Delta \widetilde{S}\|_{\mathrm{row},\max}  
=  
\max_{i=1,2,3} \left(\sum_{j=1}^{3} |\Delta \widetilde{S}_{ij}|^2\right)^{1/2}.  
\end{equation}
This criterion decouples the local accuracy requirement of the propagator from the history-dependent incoming state, while still ensuring control over the largest possible component-wise error incurred when the operator acts on any normalized state. In practice, the defect operator is computed via step-doubling. For a candidate cell of length \(\Delta x\), we first compute the coarse propagator over the full cell, denoted \(\widetilde S^\odot_{\rm c}(\Delta x)\); then, we compute the fine propagator by splitting the same cell into two equal subcells:  
\[
\widetilde S^\odot_{\rm f}(\Delta x)=\widetilde S^\odot_{\rm R}(\Delta x/2)\,\widetilde S^\odot_{\rm L}(\Delta x/2).
\]  
Before computing their difference, we remove the common global phase between the two propagators. Specifically, we select a phase factor \(\alpha\) that minimizes  
\[
\left\|\widetilde S^\odot_{\rm c}-\alpha \widetilde S^\odot_{\rm f}\right\|_{\rm F}.
\]  
The optimal phase is given by  
\[
\alpha = \frac{ \left[\operatorname{Tr}\left(\widetilde S_{\rm c}^{\odot\dagger}\widetilde S^\odot_{\rm f}\right)\right]^* }{ \left|\operatorname{Tr}\left(\widetilde S_{\rm c}^{\odot\dagger}\widetilde S^\odot_{\rm f}\right)\right| }.
\]  
The local defect operator employed in the tolerance test is thus defined as  
\[
\Delta \widetilde S = \widetilde S^\odot_{\rm c}-\alpha \widetilde S^\odot_{\rm f}.
\]  
For a fourth-order integration step, the coarse-step error is estimated from the step-doubling difference as  
\[
\eta = \frac{16}{15} \|\Delta \widetilde S\|_{\mathrm{row},\max}.
\]  
A cell is accepted if \(\eta \leq \epsilon_{\rm pred}\), where \(\epsilon_{\rm pred}\) denotes the prescribed predictor tolerance.

\subsubsection{The cheap predictor: the $SU(2)$ closed-form solution}
\label{sec:su2_solver}
Since the mesh need only resolve the dominant variation of the local evolution operator along the integration path, the two-flavor predictor requires only a single algebraic ingredient: the closed-form exponential of a generic \(2\times2\) Hermitian matrix. Let  
\begin{equation}  
T_{i,k,n}  
=  
\begin{pmatrix}  
a & z\\  
z^* & d  
\end{pmatrix}  
=  
\tau I_2+\boldsymbol r\cdot\boldsymbol\sigma,  
\end{equation}  
where  
\begin{equation}  
\tau=\frac{a+d}{2},  
\qquad  
\boldsymbol r  
=  
\left(  
\mathrm{Re}\,z,\,  
-\mathrm{Im}\,z,\,  
\frac{a-d}{2}  
\right),  
\qquad  
\boldsymbol\sigma=(\sigma_x,\sigma_y,\sigma_z).  
\end{equation}  
Noting that  
\begin{equation}  
\left(\boldsymbol r\cdot\boldsymbol\sigma\right)^2  
=  
\omega^2 I_2,  
\qquad  
\omega=|\boldsymbol r|  
=  
\sqrt{\left(\frac{a-d}{2}\right)^2+|z|^2},  
\end{equation}  
the matrix exponential evaluates to  
\begin{equation}  
\exp(-i\Delta x\, T_{i,k,n})  
=  
e^{-i\Delta x\,\tau}  
\left[  
\cos(\Delta x\, \omega)\,I_2  
-  
i\frac{\sin(\Delta x\, \omega)}{\omega}  
\boldsymbol r\cdot\boldsymbol\sigma  
\right],  
\label{eq:su2_closed_form}  
\end{equation}  
with the regularized limit \(\sin(\Delta x\cdot \omega)/\omega \to \Delta x\) as \(\omega \to 0\).  

In the present adaptive-mesh construction, the propagators \(\widetilde S_{\rm c}\), \(\widetilde S_{\rm L}\), and \(\widetilde S_{\rm R}\) appearing in the step-doubling test are all predictor propagators expressed in the CF4 two-stage form. For each computational cell, the two stage-to-stage transition factors (stages 1→2) are computed by applying Eq.~\eqref{eq:su2_closed_form} to the corresponding active \(2\times2\) block.

\subsubsection{Dyadic refinement and binary mesh storage}
With the tolerance criterion and the \(SU(2)\) closed-form evaluator defined, we can construct the predictor propagators used in the step-doubling test. The coarse predictor \(\widetilde{S}_{\rm c}(\Delta x)\) is assembled from the two CF4 stage factors over the full cell, whereas the fine predictor \(\widetilde{S}_{\rm f}(\Delta x)\) is assembled from the four CF4 stage factors obtained by applying the integrator to the two half-cells.  

For a trajectory with impact parameter \(b\), let \(x_0\) denote the starting point and \(L\) the total path length. The initial mesh consists of a single root cell spanning the entire interval \([x_0,\, x_0 + L]\). Each cell is uniquely identified by a refinement level \(\ell\) and an integer left index \(j\). The physical interval associated with the cell \((\ell,j)\) is given by  
\begin{equation}  
h_\ell = \frac{L}{2^\ell}, \qquad [x_{\rm left},\, x_{\rm right}] = [x_0 + j h_\ell,\, x_0 + (j+1) h_\ell],  
\label{eq:dyadic_cell}  
\end{equation}  
so that the root cell corresponds to \((\ell,j) = (0,0)\).  

Refinement proceeds leaf-by-leaf. In each round, the predictor error is evaluated only for the currently active leaves. A cell satisfying the tolerance criterion defined above is accepted as a final leaf of the adaptive mesh and excluded from further refinement. Conversely, if a cell fails the criterion, it is subdivided into its two dyadic children:  
\begin{equation}  
(\ell,j) \longrightarrow (\ell+1,\, 2j), \qquad (\ell+1,\, 2j+1).  
\label{eq:dyadic_split}  
\end{equation}  
Thus, each refinement round operates exclusively on the active frontier of the binary tree, avoiding full reconstruction of the mesh at every step.  

The energy range \([0.1,\, 16]\) MeV for \(^8\)B neutrinos is partitioned into several bins; for example, \([0.1,\, 4.0,\, 8.0,\, 12.0,\, 16.0]\) MeV. For a given energy bin, the cell error employed in the accept/reject test is defined as the maximum predictor error across all energy grid points within that bin:  
\begin{equation}  
\eta_{\ell,j}^{\rm bin} = \max_{E_k\in {\rm bin}} \eta_{\ell,j}(E_k).  
\end{equation}  
A cell is accepted if and only if  
\begin{equation}  
\eta_{\ell,j}^{\rm bin} \leq \epsilon_{\rm pred}.  
\end{equation}  
Otherwise, it is bisected according to Eq.~\eqref{eq:dyadic_split}.  
This dyadic representation is both compact and reproducible. Rather than storing floating-point boundary values generated through iterative subdivision, each accepted segment is encoded solely by its trajectory index and the associated integer pair \((\ell,j)\). The corresponding physical boundaries are reconstructed via Eq.~\eqref{eq:dyadic_cell} only upon materialization of the final propagation arrays. At each refinement iteration, predictor evaluations, which constitute the dominant computational cost, are parallelized over both the set of active leaf cells and the energy grid points within each bin. Subsequently, the energy dimension is reduced to the bin-wise maximum error \(\eta_{\ell,j}^{\rm bin}\) for each cell, which governs the accept/split decision in the next step.

\section{Unsupervised machine learning for path-space compression} 

\subsection{Path representation as spherical point clouds}
The 1D Earth model is radially symmetric; thus, for a fixed detector location, the matter profile along the neutrino path depends solely on the nadir angle $\eta$ (where, during nighttime, $\eta \in [0^\circ, 90^\circ]$). In contrast, the 3D Earth model necessitates an additional angular coordinate, the azimuthal angle $az$, to fully specify Earth-crossing directions. In practice, we compute the time-dependent geometric quantities:   
\[
[\mathcal{R}_{\odot d}(t_i),\ \eta(t_i),\ az(t_i),\ \texttt{is\_night}(t_i)]
\]  
at each discrete time point $t_i$, using a temporal resolution of $1~\mathrm{min}$. Here, $\eta(t_i)$ and $az(t_i)$ define the direction from the detector to the solar center and thereby determine the central ray of the solar-neutrino path ensemble. The quantity $\mathcal{R}_{\odot d}(t_i)$ denotes the instantaneous distance between the detector and the solar center. When required, the corresponding geometric flux factor is given by $\Phi(t_i) \propto \mathcal{R}_{\odot d}^{-2}(t_i)$. The Boolean variable $\texttt{is\_night}(t_i)$ serves to identify nighttime intervals, enabling selection of only those samples for which the neutrino trajectory traverses the Earth. As a result, the full set comprises $262{,}167$ distinct nighttime Earth-crossing directions, a number that renders exhaustive computational treatment infeasible. Consequently, a principled strategy must be introduced to select representative trajectories and appropriately bin the remaining ones.

Owing to the periodic nature of the angular coordinates $(\eta, az)$, it is natural to map them onto a unit vector in the East–North–Up (ENU) coordinate system:  
\begin{equation}
\mathbf{u}_i = \big(\cos(\mathrm{alt}_i)\sin(\mathrm{az}_i),\ \cos(\mathrm{alt}_i)\cos(\mathrm{az}_i),\ \sin(\mathrm{alt}_i)\big),
\label{center_unit_vec}
\end{equation}  
where $\mathrm{alt}_i = \eta_i - \pi/2$. Each such unit vector $\mathbf{u}_i$ represents the central direction of a nighttime Earth-crossing trajectory. When accounting for the finite spatial extent of the $^8\mathrm{B}$ neutrino production region in the Sun, this central direction should be interpreted not as a single deterministic path, but rather as the centroid of a spherical point cloud in path space—reflecting the intrinsic angular spread of the neutrino ensemble. As demonstrated in~\ref{Gaussian_cloud}, the finite-source angular cloud surrounding each time-dependent central direction is isotropic by construction. It is numerically well approximated, within the local tangent plane, by a two-dimensional Gaussian distribution. Consequently, each time sample can be compactly represented by the pair  
\begin{equation}  
\mathcal{C}_i = \left(\mathbf{u}_i,\sigma_{\rm tan}(t_i)\right),  
\end{equation}  
where $\mathbf{u}_i$ denotes the central Earth-crossing direction, and $\sigma_{\rm tan}(t_i)$ quantifies the transverse angular width of the $^8\mathrm{B}$ finite-source cloud.

\subsection{A spherical $W_2$-inspired surrogate distance}
\label{W2_distance}
For point clouds residing in the Euclidean plane, a natural and mathematically principled way to characterize their spatial distribution is the quadratic optimal transport distance, specifically, the Wasserstein-2 distance. In particular, when the point clouds are modeled as Gaussian distributions, the squared Wasserstein-2 distance admits a closed-form expression~\cite{Gelbrich:1990}:  
\begin{equation}
W_2^2
=
\|\boldsymbol{\mu}_i-\boldsymbol{\mu}_j\|^2
+
\mathrm{Tr}\left[
\Sigma_i+\Sigma_j
-
2\left(\Sigma_j^{1/2}\Sigma_i\Sigma_j^{1/2}\right)^{1/2}
\right],
\end{equation}  
where the two planar Gaussian clouds are denoted by $\mathcal{N}(\boldsymbol{\mu}_i,\Sigma_i)$ and $\mathcal{N}(\boldsymbol{\mu}_j,\Sigma_j)$, and $\Sigma_i$, $\Sigma_j$ are their respective $2 \times 2$ covariance matrices. For the isotropic Gaussian clouds considered herein—i.e., those satisfying $\Sigma_i = \sigma_i^2 I_2$ and $\Sigma_j = \sigma_j^2 I_2$—this expression simplifies to  
\begin{equation}
W_2^2
=
\|\boldsymbol{\mu}_i-\boldsymbol{\mu}_j\|^2
+
2(\sigma_i-\sigma_j)^2 .
\label{eq:planar_iso_gaussian_w2}
\end{equation}  
Intuitively, Eq.~\eqref{eq:planar_iso_gaussian_w2} decomposes the optimal transport cost into two geometrically interpretable components: a rigid translation that aligns the centers of the clouds, and an isotropic scaling that adjusts their spreads about those centers.

In the present problem, our point clouds reside on the unit sphere rather than in a Euclidean plane. Because the sphere admits no global translation operation, displacing the cloud center along its geodesic does not uniquely determine the shortest displacement for all off-center points within the cloud. However, the angular extent of the $^8\mathrm{B}$ finite-source cloud is exceedingly small. For a production radius $r_{\rm prod}\simeq 0.16R_\odot$, where $R_\odot$ denotes the solar radius, the maximum angular radius, viewed from Earth, is approximately  
\begin{equation} 
q_{\rm max} \simeq \arctan\left(\frac{0.16R_\odot}{1~\mathrm{AU}}\right) \simeq 7.4\times 10^{-4}\ {\rm rad}. 
\end{equation}  
Given that the finite-source angular cloud is highly localized, each spherical cloud can be accurately represented using local tangent coordinates centered on its mean direction. When the centers of two such clouds are also proximal on the sphere, the pairwise transport from $\mathcal C_i$ to $\mathcal C_j$ exhibits the same leading-order structure as the affine Gaussian-to-Gaussian transport in Euclidean space: the cloud center is relocated from $\mathbf u_i$ to $\mathbf u_j$; the tangent-space fluctuations are mapped via parallel transport along the geodesic connecting the centers; and the isotropic width is rescaled from $\sigma_{{\rm tan}, i}$ to $\sigma_{{\rm tan},j}$.  

For finite center-to-center separations, this affine structure should not be interpreted as an exact small-cloud expansion of the intrinsic spherical optimal transport distance. Nevertheless, within the clustering objective, the dissimilarity measure is dominated by the geodesic distance between centers in this regime, since $\sigma_{\rm tan} = O(q_{\rm max})$; the contribution from cloud width constitutes only an $O(q_{\rm max}^2)$ correction to the leading-order geodesic separation. This width-related term becomes significant primarily when competing cluster centers exhibit geodesic separations that are comparable at the $O(q_{\rm max}^2)$ level—that is, in the local regime where the tangent-plane $W_2$ distance provides a valid approximation. We therefore adopt the Gaussian-motivated isotropic width-mismatch term as a local, finite-source refinement of an otherwise geodesic clustering dissimilarity.  

Accordingly, we define the squared pairwise dissimilarity employed in unsupervised \(k\)-medoids clustering as  
\begin{equation}  
\mathcal{D}_{ij}^2  
=  
\theta_{ij}^2  
+  
2\left(\sigma_{{\rm tan},i} - \sigma_{{\rm tan},j}\right)^2,  
\qquad  
\theta_{ij}  
=  
\arccos(\mathbf{u}_i \cdot \mathbf{u}_j).  
\label{eq:w2_inspired_kmedoids_distance}  
\end{equation}  
Although motivated by the isotropic Gaussian result, the width term in Eq.~\eqref{eq:w2_inspired_kmedoids_distance} is not restricted to Gaussian clouds. For a fixed neutrino source, the projected profiles at different Sun--detector distances form, to high numerical accuracy, an isotropic location--scale family: they share the same radial shape and
differ only in their tangent-space scale. Their planar quadratic Wasserstein distance therefore contains the same width contribution, $2(\sigma_{{\rm tan}, i}-\sigma_{{\rm tan},j})^2$, independently of the detailed radial profile. Among the eight BS05(OP) source profiles, we verify that all except ${}^{13}\mathrm{N}$ are approximately Gaussian, whereas all eight sources, including ${}^{13}\mathrm{N}$, exhibit scale-family behavior.

Under this formulation, the dissimilarity jointly penalizes both the geodesic separation between cluster centers and the mismatch in tangent-space widths. It serves as a task-specific objective for unsupervised \(k\)-medoids clustering—rather than an exact analytical expression for the intrinsic spherical Wasserstein distance between the underlying point clouds.

\subsection{$K$-means initialization and $K$-medoids clustering}
\label{sec:kmedoids}
For the pairwise dissimilarity defined in Eq.~\eqref{eq:w2_inspired_kmedoids_distance}, our objective is to compress the large collection of nighttime path-space clouds into $K$ representative trajectories. Since the computation of MSW evolution on the Earth’s side relies on physically realized trajectories, we select an actual time sample from each cluster, as opposed to constructing an artificial barycenter, to serve as the final representative trajectory. Consequently, we adopt a $K$-medoids formulation~\cite{Kaufman:1987}, wherein each cluster is represented by one of its constituent samples.

Given a set of selected nighttime samples $\{\mathcal{C}_i\}_{i=1}^{N_{\rm night}}$, let  
\[
\mathcal{M} = \{m_1,\ldots,m_K\}
\]  
denote the set of medoid indices, where each $m_k$ corresponds to the index of an actual time sample. We minimize the unweighted $K$-medoids objective  
\begin{equation}
\mathcal{L}(\mathcal{M}) = \sum_{i=1}^{N_{\rm night}} \min_{1 \leq k \leq K} \mathcal{D}_{i m_k}^2,
\label{eq:kmedoids_objective}
\end{equation}  
where $\mathcal{D}_{i m_k}^2$ denotes the squared dissimilarity defined in Eq.~\eqref{eq:w2_inspired_kmedoids_distance}. For fixed medoids, each sample is assigned to its nearest representative according to  
\begin{equation}
\ell_i = \arg\min_{1\le k\le K} \mathcal{D}_{i m_k}^2 .
\end{equation}  
Given fixed cluster assignments \(C_k = \{i: \ell_i = k\}\), the medoid of each cluster is updated by selecting the actual sample within that cluster which minimizes the sum of squared dissimilarities to all other samples in the same cluster:  
\begin{equation}
m_k = \arg\min_{j\in C_k} \sum_{i\in C_k} \mathcal{D}_{ij}^2 .
\label{eq:kmedoids_medoid_update}
\end{equation}  
Unsupervised \(K\)-medoids clustering is then performed by iteratively alternating between these two steps: the assignment step minimizes the objective function in Eq.~\eqref{eq:kmedoids_objective} with respect to the cluster labels, while the medoid-update step applies Eq.~\eqref{eq:kmedoids_medoid_update} under fixed cluster assignments. The algorithm terminates when the medoid indices, cluster assignments, or objective value converge—i.e., no longer change across successive iterations.  

The above alternating optimization procedure fully defines the \(K\)-medoids clustering algorithm once initial medoids are specified. A standard initialization strategy involves randomly selecting \(K\) samples from the full nighttime dataset. However, both the quality of the final solution and the convergence speed are highly sensitive to the choice of initial medoids. To mitigate this sensitivity, we introduce a preliminary, \(K\)-means–inspired~\cite{Lloyd:1982} prototype stage—designed exclusively to coarsely refine the random initialization before executing the final \(K\)-medoids refinement.

This prototype stage adopts the same assignment rule as \(K\)-medoids, based on squared dissimilarity \(\mathcal{D}_{ij}^2\). Crucially, unlike the medoid-update step in Eq.~\eqref{eq:kmedoids_medoid_update}, the representative at this stage is permitted to be a continuous prototype, that is, a point in the dissimilarity space, rather than being constrained to coincide with an actual observed sample.

For fixed assignments, the prototype is updated via a fast chordal-distance mean: the unit vectors corresponding to assigned samples are averaged in the ambient \(\mathbb{R}^3\) space and then renormalized to lie on \(\mathbb{S}^2\).

Upon convergence of the prototype stage, each prototype cluster is converted into an actual time sample by applying the medoid update in Eq.~\eqref{eq:kmedoids_medoid_update}. The resulting medoids serve as the initial representatives for the final \(K\)-medoids refinement. Consequently, the final output comprises \(K\) physically realizable representative trajectories along with their associated cluster weights.

\begin{figure*}[t]
    \centering
    \includegraphics[width=0.98\textwidth]{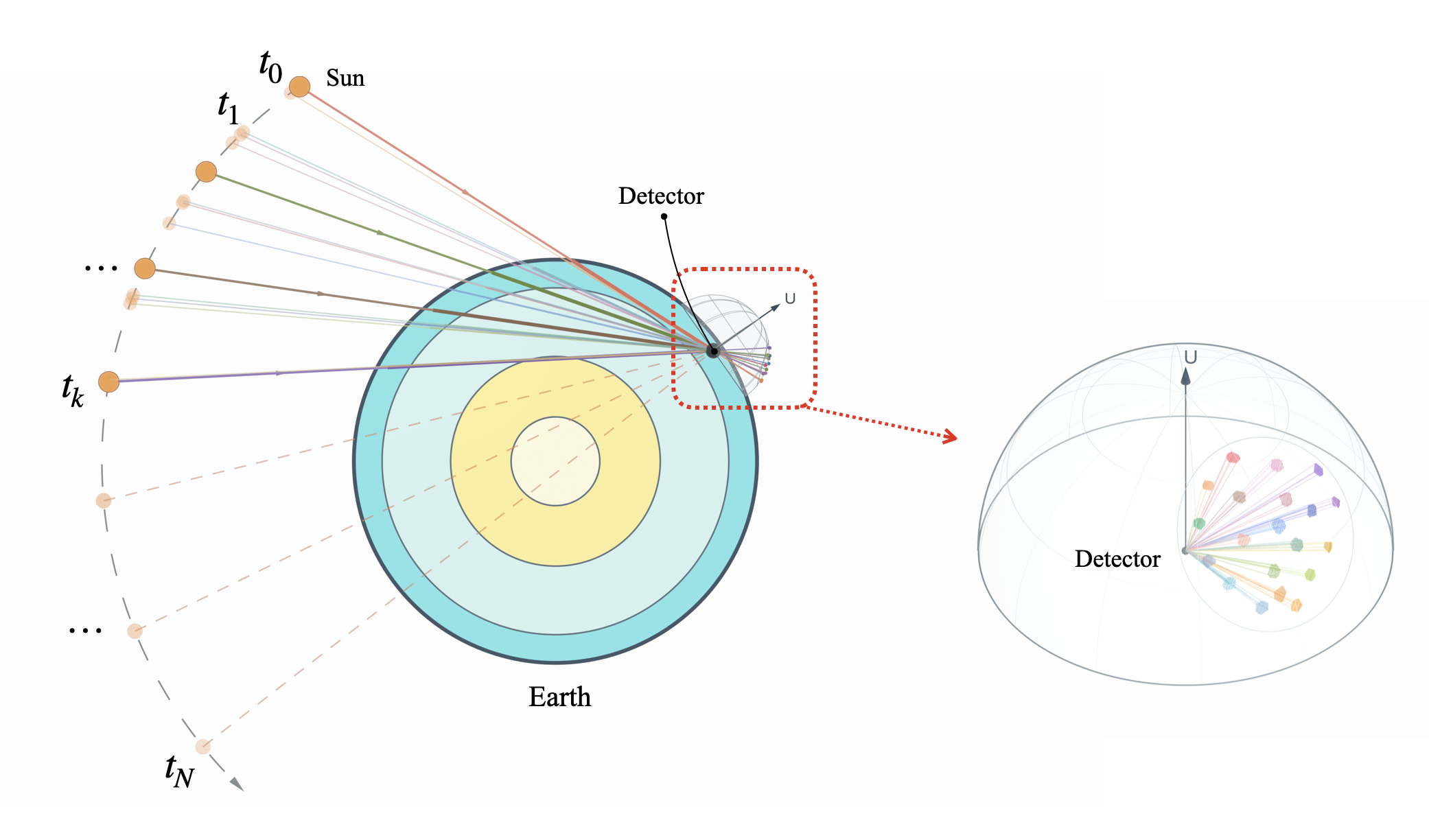}
    \caption{
    Geometric origin and detector-centered path-space representation of nighttime solar neutrino trajectories.  
Left panel: Distinct nighttime positions of the Sun give rise to distinct Earth-crossing neutrino trajectories, all converging at the underground detector. The inset centered on the detector depicts the associated local directional space, where $\hat{\mathbf{U}}$ denotes the local upward direction. The dashed callout highlights the region magnified in the left panel.  
Right panel: Magnified unit-hemisphere representation of local neutrino propagation directions. Each ray originates from the detector-centered origin and terminates at a sampled direction on the unit sphere; endpoint samples belonging to the same trajectory cluster are rendered in identical color. For visual clarity, only a representative subset of clusters drawn from the full ensemble of nighttime trajectories is displayed.
    }
    \label{fig:path-clustering}
\end{figure*}

\subsection{Representative-path reweighting and solar--Earth matching}
\label{sec:solar_earth_matching}
We now describe how the solar-side and Earth-side calculations are combined after selecting representative Earth trajectories.  

Regarding the solar-side output, consider an infinitesimal neutrino-production element of the selected solar component. Its initial state in the vacuum mass basis is denoted by \(|\nu_e\rangle_{\text{vac}}\). Upon forward propagation from the production point to the solar surface, this state evolves into  
\begin{equation}  
|\nu_e\rangle^\odot = \prod_{n=1}^N \widetilde S_n |\nu_e\rangle_{\text{vac}} \ ,\qquad   
|\nu_e\rangle^\odot =   
\begin{pmatrix}  
a_1\\a_2\\a_3  
\end{pmatrix},  
\end{equation}  
where \(\widetilde S_n\) denotes the CF4 evolution operator for the \(n\)th solar layer step, as defined in Eq.~\eqref{eq:CF4_2stage_OS}.  

On the Earth side, we consider the detector observable \({}_{\rm vac}\langle \nu_e|\) propagating backward along the Earth-crossing trajectory. For a given Earth-crossing trajectory \(\gamma\), we write  
\begin{equation}  
{}^\oplus\!\langle \nu_e| = {}_{\rm vac}\langle \nu_e|\prod_{n=N}^1 S^\oplus_n \ ,\qquad   
{}^\oplus\!\langle \nu_e| =   
\begin{pmatrix}  
r_1,&r_2,&r_3  
\end{pmatrix},  
\end{equation}  
where \(S^\oplus_n\) denotes the evolution operator for the \(n\)th Earth layer step, introduced in Eq.~\eqref{eq:strang_layer}.  

Between the solar surface and the Earth entry point, the mass eigenstates accumulate vacuum phases. After factoring out an irrelevant overall phase, we define the relative vacuum phase matrix as  
\begin{equation}  
\Phi_{\odot\oplus}(E,t)  
=  
\mathrm{diag}  
\left(  
1,\,  
e^{-i\phi_{21}(E,t)},\,  
e^{-i\phi_{31}(E,t)}  
\right),  
\qquad  
\phi_{i1}(E,t)  
=  
\frac{\Delta m^2_{i1}\,\mathcal{R}_{\odot d}(t)}{2E},  
\label{eq:sun_earth_vacuum_phase}  
\end{equation}  
where \(\mathcal{R}_{\odot d}(t)\) is the distance between the detector and the Sun’s center at time \(t\). Therefore, for this source element and the corresponding Earth trajectory, the survival amplitude is given by  
\begin{equation}
\mathcal A_{ee}(E,t;\gamma)
=
{}^\oplus\!\langle \nu_e|\Phi_{\odot\oplus}(E,t)|\nu_e\rangle^\odot
=
\sum_{i=1}^{3}
r_i(E;\gamma)\,
a_i(E)\,
e^{-i\phi_i(E,t)} ,
\label{eq:single_source_survival_amplitude}
\end{equation}
where one may conventionally set $\phi_1 = 0$, $\phi_2 = \phi_{21}$, and $\phi_3 = \phi_{31}$. The associated survival probability reads  
\begin{align}
P_{ee}(E,t;\gamma)
&=\left|\sum_{i=1}^{3}r_i a_i e^{-i\phi_i}\right|^2
\nonumber\\
&=\sum_{i=1}^{3}|r_i|^2 |a_i|^2+
2\,\mathrm{Re}\sum_{i<j}r_i r_j^*a_i a_j^*e^{-i(\phi_i-\phi_j)} .
\label{eq:single_source_survival_probability}
\end{align}
In conventional analyses, the rapidly oscillating vacuum phases accumulated along the Sun–Earth baseline are not explicitly retained. Upon averaging over the neutrino production region, experimental energy resolution, and time exposure, the interference terms $2\,\mathrm{Re}\sum_{i<j}r_i r_j^*a_i a_j^*e^{-i(\phi_i-\phi_j)}$ are effectively suppressed.  
Nonetheless, it remains advantageous to retain the solar neutrino output at the amplitude level, expressed in density-matrix form, before performing this phase-averaging limit. For a single source element, we define the solar density matrix  
\begin{equation}
M^\odot=\begin{pmatrix}
|a_1|^2& a_1a_2^*&a_1a_3^*\\
a_2a_1^*&|a_2|^2&a_2a_3^*\\
a_3a_1^*&a_3a_2^* &|a_3|^2
\end{pmatrix},
\end{equation}
in terms of which Eq.~\eqref{eq:single_source_survival_probability} may be recast as  
\begin{equation}
P_{ee}(E,t;\gamma)
=
\sum_i |r_i|^2 M^\odot_{ii}
+
\sum_{i\neq j}
r_i r_j^*
e^{-i(\phi_i-\phi_j)}
M^\odot_{ij}.
\label{eq:survival_probability_with_M}
\end{equation}
The finite-source averaging on the solar side can be carried out directly at the level of the matrix elements \(M^\odot_{ij}\). Denoting each source element by \(\alpha\) and assigning it a weight \(w_\alpha\), we define  
\begin{equation}  
M^{\odot,\alpha}_{ij} = a_i^\alpha a_j^{\alpha *},  
\qquad  
\overline M^\odot_{ij} = \sum_{\alpha} w_\alpha M^{\odot,\alpha}_{ij}.  
\label{eq:source_averaged_M}  
\end{equation}  
Substituting this averaged matrix into Eq.~\eqref{eq:survival_probability_with_M}, the survival probability becomes  
\begin{equation}  
P_{ee}^{N}(E,t;\gamma) = \sum_i |r_i|^2 \overline M^\odot_{ii} + \sum_{i\neq j} r_i r_j^* e^{-i(\phi_i-\phi_j)} \overline M^\odot_{ij}.  
\label{eq:source_averaged_survival_probability}  
\end{equation}  
This expression corresponds precisely to the quantity accumulated during solar-side backward propagation, as introduced in Sec.~\ref{sec:back_prop}: rather than storing the complex amplitudes from each neutrino production point separately, the algorithm constructs the weighted matrix elements \(\overline M^\odot_{ij}\) directly along each solar impact-parameter chord.  

Once the solar-side source averaging is complete, the only remaining trajectory dependence in Eq.~\eqref{eq:source_averaged_survival_probability} resides in the Earth-side row amplitudes \(r_i(E;\gamma)\). The unsupervised machine-learning clustering procedure outlined in Sec.~\ref{sec:kmedoids} partitions the original set of nighttime trajectories into clusters \(C_k\), each represented by a medoid trajectory \(\gamma_{m_k}\).
Within each cluster, we approximate the Earth-side propagation result for all trajectories by that of the corresponding medoid:  
\begin{equation}  
r_i(E;\gamma_a) \simeq r_i(E;\gamma_{m_k}), \qquad \gamma_a \in C_k.  
\label{eq:cluster_medoid_row_approx}  
\end{equation}  
Equivalently, for the bilinear combinations appearing in the survival probability, we have  
\begin{equation}  
\sum_{\gamma_a \in C_k} w_a\, r_i(E;\gamma_a)\, r_j^*(E;\gamma_a) \simeq \Omega_k\, r_i(E;\gamma_{m_k})\, r_j^*(E;\gamma_{m_k}), \qquad \Omega_k = \sum_{\gamma_a \in C_k} w_a.  
\end{equation}  
In the present implementation, all nighttime samples are assigned equal weights; thus, the weight per trajectory and the total weight of cluster \(C_k\) are given by  
\begin{equation}  
w_a = \frac{1}{N_{\rm night}}, \qquad \Omega_k = \frac{|C_k|}{N_{\rm night}}.  
\label{eq:cluster_equal_weight}  
\end{equation}  
The compressed nighttime survival probability can then be expressed as  
\begin{align}  
\overline P_{ee}^{N}(E) \simeq \sum_{k=1}^{K} \Omega_k \Bigg[ \sum_i \big|r_i^{(m_k)}(E)\big|^2 \overline M^\odot_{ii}(E) + \sum_{i\neq j} r_i^{(m_k)}(E)\, r_j^{(m_k)*}(E)\, e^{-i(\phi_i - \phi_j)} \overline M^\odot_{ij}(E) \Bigg].  
\label{eq:cluster_bilinear_medoid_approx}  
\end{align}

\subsection{Propagation of parametric uncertainties}
\label{sec:uncertainty_propagation}
The uncertainties associated with the neutrino oscillation parameters are propagated using a quasi-Monte Carlo (QMC) ensemble based on a scrambled Sobol sequence~\cite{Sobol1967, Virtanen2020}. We generate $N_{\mathrm{QMC}} = 2^9 = 512$ samples in the four-dimensional parameter space  
\begin{equation}    
\left\{    
\sin^2\theta_{12},    
\sin^2\theta_{13},    
\Delta m^2_{21},    
\Delta m^2_{32}    
\right\}.
\end{equation}  
For each parameter, the corresponding Sobol coordinate is transformed via the inverse cumulative distribution function of a two-piece normal distribution—enabling asymmetric treatment of upper and lower uncertainties. The resulting distributions are truncated at the quoted $3\sigma$ boundaries. The same Sobol sample set is employed concurrently in both solar- and Earth-side calculations, thereby preserving the sample-wise correlation between the final observables and the shared oscillation parameters. Input-parameter correlations among the oscillation parameters are neglected in the present analysis.  

On the solar side, the parametric uncertainty propagation may be performed using either the SU(2)+1 solver or the full three-flavor solver. As a representative case, we validated this consistency at the China Jinping Underground Laboratory (CJPL) by applying the identical $N_{\mathrm{QMC}} = 512$ Sobol sample set to both solvers. Over the full energy range of $0.1$–$16~\mathrm{MeV}$, the maximum relative differences in the half-widths of the 16th–84th percentile intervals were found to be $0.67\%$ for $P_{ee}^{D}(E)$ and $0.89\%$ for $P_{ee}^{N}(E)$. The corresponding 99th-percentile differences across the energy grid were $0.64\%$ and $0.65\%$, respectively. For spectrum-integrated observables, the relative differences in the half-widths remained below $0.5\%$ for both the day- and night-time electron-neutrino fluxes, and below $0.2\%$ for the day–night asymmetries.

The uncertainty in the normalized $^{8}\mathrm{B}$ spectral shape is treated independently. We perform piecewise linear interpolation between the central spectral template and the quoted $\pm3\sigma$ templates, using a standard-normal nuisance parameter truncated at $\pm3\sigma$. From this procedure, we generate $N_{\mathrm{spec}} = 5000$ independent spectral realizations. Each realization is renormalized to unit integral, ensuring that this uncertainty affects only the spectral shape—not the adopted total $^{8}\mathrm{B}$ flux normalization. Oscillation-parameter uncertainties and spectral-shape uncertainties are assumed to be statistically independent; they are combined via product sampling when computing integrated fluxes and day–night asymmetries.

\section{Results for CJPL}
\label{sec:cjpl_result}
We begin by presenting the site-specific predictions for CJPL. The central results are derived using a high-precision ephemeris-based geometric model, a three-dimensional terrestrial electron-density model, a full three-flavor treatment of solar neutrino evolution, and the coherent solar–vacuum–Earth matching formalism described in Sec.~\ref{sec:solar_earth_matching}. The uncertainty bands shown below are obtained from the ensemble method detailed in Sec.~\ref{sec:uncertainty_propagation}. The oscillation parameters are taken from the normal-ordering entry labeled ``Ref.~[193] w SK-ATM \& IC24'' in Table~14.7 of the 2025 Particle Data Group review~\cite{ParticleDataGroup:2024}:
$\sin^2\theta_{12}=0.308^{+0.012}_{-0.011}$,
$\sin^2\theta_{13}=0.02215^{+0.00056}_{-0.00058}$,
$\Delta m^2_{21}=7.49^{+0.19}_{-0.20}\times10^{-5}\,\mathrm{eV}^2$,
and
$\Delta m^2_{32}=2.438^{+0.021}_{-0.019}\times10^{-3}\,\mathrm{eV}^2$.
All annual averages reported are evaluated using one-minute samples over the UTC interval from 2025-12-31 00:00:00 to 2026-12-31 00:00:00.

\begin{figure*}[!htbp]    
\centering    
\begin{subfigure}[t]{0.49\textwidth}        
\centering        
\includegraphics[width=\linewidth]{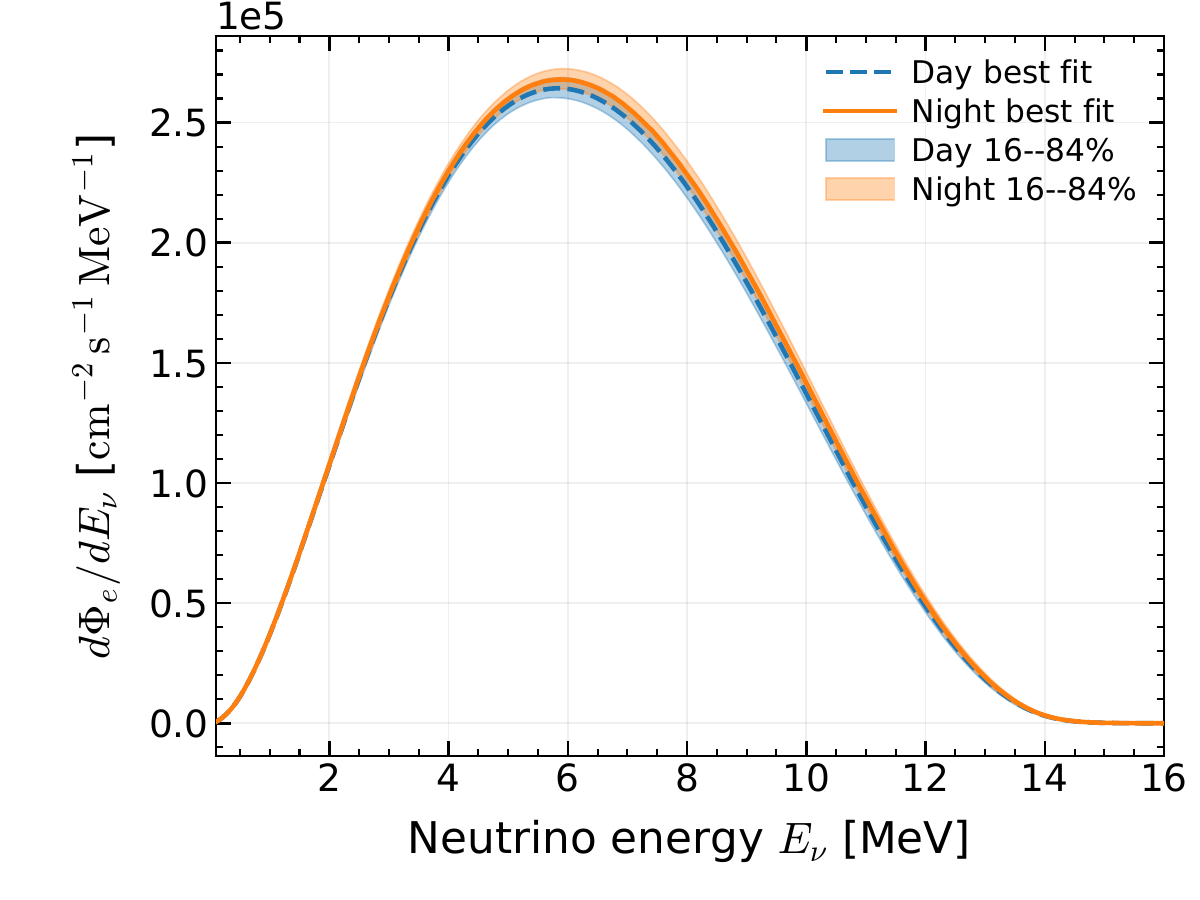}        
\caption{        
Predicted daytime and nighttime ${}^{8}\mathrm{B}$ electron-neutrino flux spectra at CJPL.        
The solid and dashed curves represent the best-fit predictions for daytime and nighttime, respectively;        
the shaded regions indicate the corresponding 16th–84th percentile uncertainty intervals.        
}        
\label{fig:cjpl_flux_spectrum}    
\end{subfigure}    
\hfill    
\begin{subfigure}[t]{0.49\textwidth}        
\centering        
\includegraphics[width=\linewidth]{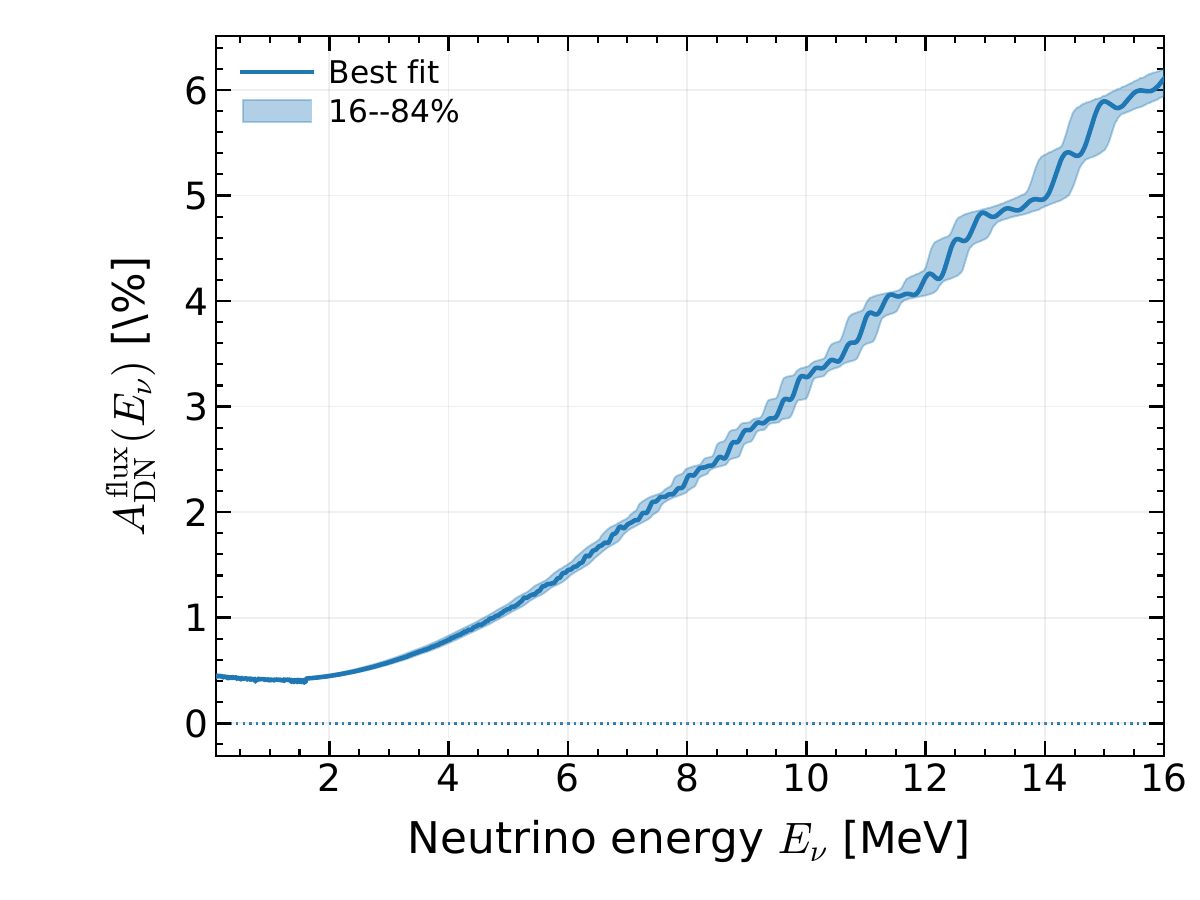}        
\caption{        
Energy-dependent day–night flux asymmetry at CJPL.        
The curve shows the best-fit prediction, and the shaded region denotes the 16th–84th percentile uncertainty interval.        
}        
\label{fig:cjpl_adn_flux}    
\end{subfigure}
    \caption{    
Site-specific ${}^{8}\mathrm{B}$ solar neutrino predictions for CJPL.        
Panel~\subref{fig:cjpl_flux_spectrum} displays the predicted daytime and nighttime electron-neutrino flux spectra, incorporating combined uncertainties from oscillation parameters and spectral shape.        
Panel~\subref{fig:cjpl_adn_flux} presents the corresponding energy-dependent day–night flux asymmetry,        
$A_{\mathrm{DN}}^{\mathrm{flux}}(E_\nu)$.    
}    
\label{fig:cjpl_flux_and_adn}
\end{figure*}
For $X\in\{D, N\}$, the annual daytime- or nighttime-averaged differential electron-neutrino flux shown in  
Fig.~\ref{fig:cjpl_flux_and_adn} is defined as  
\begin{equation}  
\frac{\mathrm{d}\Phi_e^{X}}{\mathrm{d}E_\nu}  
=  
\Phi_{{}^{8}\mathrm{B}}^{1\,\mathrm{AU}}\,  
\bar{g}_{X}\,  
f_{{}^{8}\mathrm{B}}(E_\nu)\,  
\overline{P}_{ee}^{X}(E_\nu),  
\label{eq:cjpl_differential_flux}  
\end{equation}  
where  
\begin{equation}  
g(t)  
=  
\left[  
\frac{1\,\mathrm{AU}}{R_{\odot d}(t)}  
\right]^2,  
\qquad  
\bar{g}_{X}  
=  
\left\langle g(t)\right\rangle_X,  
\label{eq:cjpl_gbar}  
\end{equation}  
and the effective day- and night-side survival probabilities are defined via inverse-square-distance weighting:  
\begin{equation}  
\overline{P}_{ee}^{X}(E_\nu)  
=  
\frac{  
\left\langle  
g(t)\,P_{ee}(E_\nu,t)  
\right\rangle_X  
}{  
\bar{g}_{X}  
}.  
\label{eq:cjpl_fluxweighted_pee}  
\end{equation}  
Here, $\langle\cdots\rangle_D$ and $\langle\cdots\rangle_N$ denote time averages over the daytime and nighttime intervals, respectively.  
Defining the spectrum-weighted survival probabilities as  
\begin{equation}  
I_X  
=  
\int \mathrm{d}E_\nu\,  
f_{{}^{8}\mathrm{B}}(E_\nu)  
\overline{P}_{ee}^{X}(E_\nu),  
\qquad X\in\{D,N\},  
\label{eq:cjpl_integrated_pee}  
\end{equation}  
the integrated electron-neutrino fluxes are given by  
\begin{align}  
\Phi_e^D  
&=  
\Phi_{{}^{8}\mathrm{B}}^{1\,\mathrm{AU}}  
\bar{g}_D I_D  
=  
(2.026^{+0.030}_{-0.026})  
\times10^6~  
\mathrm{cm}^{-2}\,\mathrm{s}^{-1}, \\  
\Phi_e^N  
&=  
\Phi_{{}^{8}\mathrm{B}}^{1\,\mathrm{AU}}  
\bar{g}_N I_N  
=  
(2.062^{+0.031}_{-0.026})  
\times10^6~  
\mathrm{cm}^{-2}\,\mathrm{s}^{-1}.  
\end{align}  
We distinguish between the day–night asymmetry of the electron-neutrino flux at the detector location, defined as  
\begin{equation}  
A_{\mathrm{DN}}^{\mathrm{flux}}  
=  
2\frac{\Phi_e^{N}-\Phi_e^{D}}{\Phi_e^{N}+\Phi_e^{D}},  
\label{eq:adn_flux}  
\end{equation}  
and the geometry-normalized oscillation-induced asymmetry in the survival probability, defined as  
\begin{equation}  
A_{\mathrm{DN}}^{\mathrm{osc}}  
=  
2\frac{I_N-I_D}{I_N+I_D}.  
\label{eq:adn_osc}  
\end{equation}  
For CJPL, we obtain  
\begin{equation}  
A_{\mathrm{DN}}^{\mathrm{flux}}  
=  
(1.775^{+0.079}_{-0.068})\%,  
\qquad  
A_{\mathrm{DN}}^{\mathrm{osc}}  
=  
(1.313^{+0.079}_{-0.068})\%.  
\end{equation}

The enhancement of the night-side flux arises from two contributions: Earth matter regeneration and the difference between the daytime and nighttime averages of the inverse-square Sun–detector distance factor. After explicitly accounting for the difference between $\bar{g}_D$ and $\bar{g}_N$, the residual oscillation-driven asymmetry is quantified by $A_{\mathrm{DN}}^{\mathrm{osc}}$.

The B16-GS98 value~\cite{Vinyoles:2016djt} is adopted here as a fixed reference normalization. Its associated solar-model normalization uncertainty is excluded from the reported uncertainty bands. Consequently, the quoted intervals reflect only the conditional uncertainties stemming from the oscillation parameters and the normalized ${}^{8}\mathrm{B}$ spectral shape. The total-flux normalization serves as a common multiplicative factor for both the daytime and nighttime fluxes and therefore cancels exactly in the computation of day–night asymmetries. Including this normalization uncertainty in the absolute-flux uncertainty bands would dominate the significantly smaller propagation-related uncertainties, without influencing either the predicted day–night asymmetries or their dependence on detector site.
\section{Results for Multiple Detector Sites}
We apply the same geometric, oscillatory, and uncertainty-propagation framework to a representative set of underground laboratories. Table~\ref{tab:site_flux_asymmetry} summarizes the predicted daytime and nighttime ${}^{8}\mathrm{B}$ electron-neutrino fluxes, along with the corresponding flux-level and geometry-normalized day–night asymmetries. All quantities are defined identically to those introduced in Sec.~\ref{sec:cjpl_result}. For the inverted-ordering comparison, the corresponding IO parameters are
$\sin^2\theta_{12}=0.308^{+0.012}_{-0.011}$,
$\sin^2\theta_{13}=0.02221^{+0.00056}_{-0.00056}$,
$\Delta m^2_{21}=7.49^{+0.19}_{-0.19}\times10^{-5}\,\mathrm{eV}^2$,
and
$\Delta m^2_{32}=-2.484^{+0.020}_{-0.020}\times10^{-3}\,\mathrm{eV}^2$.

Although the predicted daytime and nighttime fluxes vary only modestly across the sites, the flux-level day–night asymmetry exhibits a pronounced geographical dependence: it ranges from $0.759\%$ at SUPL to $3.397\%$ at CallioLab. In contrast, the geometry-normalized asymmetry, $A_{\mathrm{DN}}^{\mathrm{osc}}$, remains tightly constrained within the narrow interval $1.313\%$–$1.477\%$. This comparison reveals that a significant portion of the site-to-site variation in $A_{\mathrm{DN}}^{\mathrm{flux}}$ arises not solely from Earth matter regeneration, but rather from the correlation between local daytime–nighttime exposure patterns and the annual modulation of the inverse-square Sun–detector distance. In general, northern-hemisphere sites experience longer winter nights near perihelion, leading to $A_{\mathrm{DN}}^{\mathrm{flux}} > A_{\mathrm{DN}}^{\mathrm{osc}}$. Conversely, southern-hemisphere sites—ANDES and SUPL—exhibit the opposite trend, as their extended nighttime durations occur closer to aphelion.

We further assess the impact of differing astronomical models, as outlined in Sec.~\ref{sec:meeus} and Sec.~\ref{sec:astropy}. The relative differences in the final predicted fluxes are below $\mathcal{O}(10^{-6})$, while the absolute differences in the day–night asymmetries remain below $\mathcal{O}(10^{-5})$.

As for the systematic error defined in Table~\ref{tab:site_flux_asymmetry}, $\Delta_\oplus$ quantifies the correction introduced by adopting a three-dimensional Earth model, relative to the one-dimensional PREM model. Although $|\Delta_{\oplus}|$ remains modest, no larger than approximately $0.20$ percentage points, the magnitude of the day–night asymmetry itself is only on the order of $1$–$3\%$. Consequently, the three-dimensional correction amounts to several percent relative to the PREM-based prediction, reaching up to $\sim 8\%$ for the sites considered here. This result underscores that the three-dimensional terrestrial structure cannot be absorbed into a uniform normalization factor; rather, it exhibits a site-specific dependence tied to the azimuthal distribution of nighttime neutrino trajectories. In contrast, the difference in predicted asymmetry between inverted and normal neutrino mass ordering is merely $0.006$–$0.009$ percentage points for these sites, well below the current level of parametric uncertainty in the predicted day–night asymmetry.

As an additional validation, we compared the full numerical treatment of solar neutrino propagation with the three-flavor adiabatic approximation. The maximum deviation in the nighttime survival probability is $\mathcal{O}(10^{-5})$, confirming the excellent accuracy of the three-flavor adiabatic limit for standard ${}^{8}\mathrm{B}$ solar neutrinos.

Nonetheless, the full numerical approach serves as a robust benchmark and provides a foundational framework for future extensions beyond stationary, spherically symmetric solar density profiles. In particular, the local, phase-incoherent adiabatic treatment employed in the above comparison does not preserve path-dependent quantum phases and, therefore, cannot generally capture nonadiabatic transitions or parametric responses induced by three-dimensional, time-dependent density perturbations—such as those associated with solar gravity modes. The present full $3\times3$ numerical propagator thus constitutes a natural and flexible computational framework for forthcoming studies incorporating realistic three-dimensional solar models and dynamic density perturbations.

\begin{table*}[t]
\centering
\caption{
Predicted $^8$B solar electron-neutrino fluxes and day–night asymmetries  
at representative underground laboratories.  
The fluxes are expressed in units of $10^6~\mathrm{cm}^{-2}\,\mathrm{s}^{-1}$. The asymmetries are reported in percent, whereas $\Delta_{\oplus}$ and $\Delta_{\mathrm{I/N}}$ are given in percentage points.  
$A_{\rm DN}^{\rm osc}$ denotes the spectrum-weighted day–night asymmetry of the electron-neutrino survival probability, with the Sun–Earth distance-dependent flux normalization factor explicitly removed.  
$\Delta_{\oplus}$ quantifies the difference between day–night asymmetries computed under distinct Earth density models—specifically, $\Delta_{\oplus} = A^{\rm flux}_{\rm DN,3D} - A^{\rm flux}_{\rm DN,1D}$.  
$\Delta_{\rm I/N}$ quantifies the difference in asymmetries arising from the two possible neutrino mass orderings, defined as $\Delta_{\rm I/N} = A^{\rm flux}_{\rm DN, IO} - A^{\rm flux}_{\rm DN, NO}$.
}
\label{tab:site_flux_asymmetry}
\small
\setlength{\tabcolsep}{4.5pt}
\begin{tabular}{lcccccccc}
\toprule
\multirow{2}{*}{Lab}
&
\multicolumn{2}{c}{Flux prediction}
&
\multicolumn{2}{c}{Day--night asymmetry $[\%]$}
&
\multicolumn{4}{c}{Sys. shifts [pp]}
\\
\cmidrule(lr){2-3}
\cmidrule(lr){4-5}
\cmidrule(lr){6-9}
&
\makecell{$\Phi_e^{\rm D}$}
&
\makecell{$\Phi_e^{\rm N}$}
&
\makecell{$A_{\rm DN}^{\rm flux}$}
&
\makecell{$A_{\rm DN}^{\rm osc}$}
&
\makecell{$\Delta_{\oplus}$}
&
\makecell{$\Delta_{\rm I/N}$}

\\
\midrule
ANDES    &$2.036^{+0.031}_{-0.027}$  &$2.054^{+0.031}_{-0.026}$  &$0.898^{+0.083}_{-0.072}$  &$1.421^{+0.083}_{-0.072}$  &$0.017$    &$0.009$  \\
Baksan   &$2.022^{+0.030}_{-0.026}$  &$2.066^{+0.031}_{-0.027}$  &$2.157^{+0.080}_{-0.069}$	  &$1.324^{+0.080}_{-0.069}$  &$-0.023$    &$0.008$  \\
Boulby   &$2.018^{+0.030}_{-0.026}$  &$2.071^{+0.031}_{-0.027}$  &$2.622^{+0.079}_{-0.069}$	  &$1.335^{+0.079}_{-0.069}$  &$0.067$    &$0.006$  \\
CallioLab &$2.010^{+0.030}_{-0.026}$  &$2.080^{+0.031}_{-0.027}$	  &$3.397^{+0.080}_{-0.070}$  &$1.385^{+0.080}_{-0.070}$  &$0.200$    &$0.008$  \\
CJPL     &$2.026^{+0.030}_{-0.026}$  &$2.062^{+0.031}_{-0.026}$	  &$1.775^{+0.079}_{-0.068}$  &$1.313^{+0.079}_{-0.068}$  &$-0.098$    &$0.008$    \\
JUNO     &$2.027^{+0.030}_{-0.027}$  &$2.064^{+0.031}_{-0.026}$	  &$1.824^{+0.088}_{-0.076}$  &$1.477^{+0.088}_{-0.076}$  &$0.050$    &$0.009$  \\
Kamioka  &$2.024^{+0.030}_{-0.026}$  &$2.066^{+0.031}_{-0.027}$  &$2.052^{+0.084}_{-0.073}$	  &$1.407^{+0.084}_{-0.073}$  &$0.025$   &$0.008$  \\
LNGS     &$2.022^{+0.030}_{-0.026}$  &$2.067^{+0.031}_{-0.027}$	  &$2.161^{+0.082}_{-0.071}$	  &$1.353^{+0.082}_{-0.071}$  &$0.001$    &$0.007$  \\
LSC      &$2.022^{+0.030}_{-0.026}$  &$2.068^{+0.031}_{-0.027}$	  &$2.249^{+0.085}_{-0.074}$  &$1.430^{+0.085}_{-0.074}$  &$0.079$    &$0.008$  \\
LSM      &$2.022^{+0.030}_{-0.026}$  &$2.068^{+0.031}_{-0.027}$  &$2.266^{+0.082}_{-0.071}$  &$1.371^{+0.082}_{-0.071}$  &$0.035$    &$0.008$  \\
SNO      &$2.021^{+0.030}_{-0.026}$  &$2.069^{+0.031}_{-0.027}$  &$2.332^{+0.083}_{-0.072}$	  &$1.393^{+0.083}_{-0.072}$  &$0.064$    &$0.009$  \\
SUPL     &$2.038^{+0.031}_{-0.027}$  &$2.053^{+0.031}_{-0.026}$	  &$0.759^{+0.086}_{-0.075}$  &$1.437^{+0.086}_{-0.075}$  &$0.057$    &$0.008$  \\
SURF     &$2.022^{+0.030}_{-0.026}$  &$2.068^{+0.031}_{-0.027}$  &$2.258^{+0.082}_{-0.071}$  &$1.391^{+0.082}_{-0.071}$  &$0.049$    &$0.008$  \\
Yemilab  &$2.024^{+0.030}_{-0.026}$  &$2.066^{+0.031}_{-0.026}$	  &$2.074^{+0.085}_{-0.074}$  &$1.410^{+0.085}_{-0.074}$  &$0.032$    &$0.008$ & &  \\
\bottomrule
\end{tabular}
\end{table*}
\section{Summary}

Solar-neutrino experiments determine neutrino oscillation parameters by comparing measured energy spectra and time-dependent event rates against theoretical predictions of the solar neutrino flux. As the precision of solar-neutrino measurements continues to improve, a rigorously controlled prediction of standard-physics effects, integral to this comparison, becomes increasingly critical. This is especially true for upcoming joint analyses aiming to reconcile solar-neutrino and reactor-antineutrino determinations of the mixing angle $\theta_{12}$ and the squared mass difference $\Delta m_{21}^{2}$. Before any persistent discrepancy between these two experimental domains can be interpreted as potential evidence for CPT violation or other physics beyond the Standard Model, it is essential first to quantify the systematic shifts—and their associated uncertainties, arising from astronomical geometry, solar and terrestrial matter propagation, spectral inputs, and numerical approximations.

In this work, we develop a unified numerical framework and perform a quantitative assessment of these effects. We present a unified framework for predicting the site-dependent ${}^{8}\mathrm{B}$ solar electron-neutrino flux, incorporating: (i) the time-varying geometric configuration between the Sun and the detector; (ii) the finite spatial extent of the solar ${}^{8}\mathrm{B}$ production region; and (iii) Earth matter effects modeled using both one- and three-dimensional Earth density profiles. The astronomical geometry is derived from high-precision ephemerides and rigorously transformed into the local detector reference frame. The extended ${}^{8}\mathrm{B}$ production region is represented via a projected angular point cloud. Concurrently, the large ensemble of nighttime neutrino trajectories traversing the Earth is efficiently compressed using a spherical Wasserstein-inspired dissimilarity metric combined with a K-medoids clustering procedure.
For the solar-side evolution, we employ backward propagation combined with a two-stage, commutator-free, fourth-order Magnus discretization. The stage-wise matrix exponentials are computed via the Ohlsson–Snellman eigenvalue construction and the Cayley–Hamilton theorem, thereby circumventing generic matrix exponentiation. A two-flavor step-doubling predictor is used to generate an adaptive dyadic mesh, while the final propagation retains the full three-flavor evolution.  

On the Earth-side evolution, the Hamiltonian is decomposed into diagonal vacuum propagation and a rank-one matter interaction term, facilitating an efficient Strang-splitting implementation. The solar, vacuum, and Earth contributions are combined at the density-matrix level, thereby preserving all relevant quantum phases and finite-source information until the final exposure averaging. Convergence tests presented in~\ref{app:convergence} confirm that the adopted production settings constrain numerical errors in the oscillation probabilities to a level of a few parts in $10^{-4}$ or better.  

Finally, we provide site-specific flux predictions for CJPL and other underground laboratories currently engaged in or planning solar neutrino research programs, including corrections for diurnal asymmetry and for the regeneration probability distribution arising from Earth matter effects. Under the adopted fixed B16-GS98 flux normalization, using ${}^{8}\mathrm{B}$ as an example, we obtain for CJPL annual daytime and nighttime electron-neutrino fluxes of \(\Phi_e^D = (2.026^{+0.030}_{-0.026})\times10^6~{\rm cm}^{-2}{\rm s}^{-1}\) and \(\Phi_e^N = (2.062^{+0.031}_{-0.026})\times10^6~{\rm cm}^{-2}{\rm s}^{-1}\), respectively. The corresponding flux-level and geometry-normalized day--night asymmetries are \(A_{\rm DN}^{\rm flux}=(1.775^{+0.079}_{-0.068})\%\) and \(A_{\rm DN}^{\rm osc}=(1.313^{+0.079}_{-0.068})\%\). Across the detector sites considered, \(A_{\rm DN}^{\rm flux}\) ranges from \(0.759\%\) to \(3.397\%\), whereas \(A_{\rm DN}^{\rm osc}\) remains within the much narrower interval \(1.313\%\)--\(1.477\%\). This demonstrates that a substantial part of the geographical variation in the flux-level asymmetry originates from the correlation between local day--night exposure and the annual inverse-square Sun--detector distance modulation, rather than from Earth-matter regeneration alone. The use of the three-dimensional Earth model changes the predicted asymmetry by up to approximately \(0.20\) percentage points, corresponding to a relative correction of up to about \(8\%\) with respect to the PREM-based prediction. Other shifts associated with the neutrino mass ordering, the choice of astronomical geometry, and the solar adiabatic approximation remain at approximately the \(10^{-5}\)--\(10^{-4}\) level in the relevant dimensionless observables. These comparisons provide quantitative benchmarks for assessing whether future experimental discrepancies originate from incomplete standard-physics modeling or instead motivate tests of CPT violation or other new physics.

\section*{Acknowledgments}
This work was supported in part by the State Key Research
Development Program in China (Nos. 2022YFA1604700), and the National Natural Science Foundation of China under Grant No. 12441513, 12127808.

\newpage
\appendix
\renewcommand{\appendixname}{Appendix}

\renewcommand{\thesubsection}{A.\arabic{subsection}}
\renewcommand{\theequation}{A.\arabic{equation}}
\renewcommand{\thefigure}{A.\arabic{figure}}
\renewcommand{\thetable}{A.\arabic{table}}

\renewcommand{\figurename}{Figure}
\renewcommand{\tablename}{Table}

\setcounter{subsection}{0}
\setcounter{equation}{0}
\setcounter{figure}{0}
\setcounter{table}{0}
\section{Properties of the point clouds}
\label{Gaussian_cloud}
The finite-source angular distribution at each time step is represented by a set of weighted concentric rings in the local tangent plane centered on the solar direction. Because the azimuthal angle is uniformly sampled on each ring, the resulting point cloud is isotropic in the transverse plane by construction; the remaining question is whether its radial profile closely approximates that of a two-dimensional Gaussian distribution.

\begin{table}[t]
\centering
\caption{Numerical validation of the Gaussian-cloud approximation for the time-dependent $^8\mathrm{B}$ angular source distribution. Statistics are computed over all $262167$ nighttime samples. For a two-dimensional isotropic Gaussian distribution, the normalized radial variable $z = q / \sigma_{\rm tan}$ follows a Rayleigh distribution. 
$D_{\rm KS}$ is the Kolmogorov–Smirnov distance to Rayleigh.}
\label{tab:gaussian_cloud_validation}
\begin{tabular}{lccc}
\hline
Diagnostic & Gaussian expectation & Time average & Time standard deviation \\
\hline
$D_{\rm KS}$ 
& $0$
& $8.615\times 10^{-3}$
& $1.430\times 10^{-10}$
\\
$\langle q^4\rangle/\langle q^2\rangle^2$
& $2$
& $2.066456$
& $9.143\times 10^{-10}$
\\
$\langle q^6\rangle/\langle q^2\rangle^3$
& $6$
& $6.559446$
& $5.100\times 10^{-9}$
\\
\hline
\end{tabular}
\end{table}
For an isotropic Gaussian distribution in the tangent plane, the signed one-dimensional marginal variable \(x/\sigma_{\rm tan}\) follows a standard normal distribution, whereas the normalized radial variable \(z = q/\sigma_{\rm tan}\) follows a Rayleigh distribution. Table~\ref{tab:gaussian_cloud_validation} summarizes this validation across all nighttime samples. The small Kolmogorov–Smirnov (KS) distance between the empirical radial cumulative distribution function (CDF) and the theoretical Rayleigh CDF indicates that the normalized radial profile is consistent with the isotropic Gaussian-cloud model at the percent level. The fourth- and sixth-order moment ratios slightly exceed their ideal Gaussian values, suggesting a mild excess in the distributional tails; however, no significant deviation from the expected Gaussian-cloud morphology is observed.

\begin{figure}[!htbp]    
\centering    
\begin{subfigure}[t]{0.48\textwidth}        
\centering        
\includegraphics[width=\textwidth]{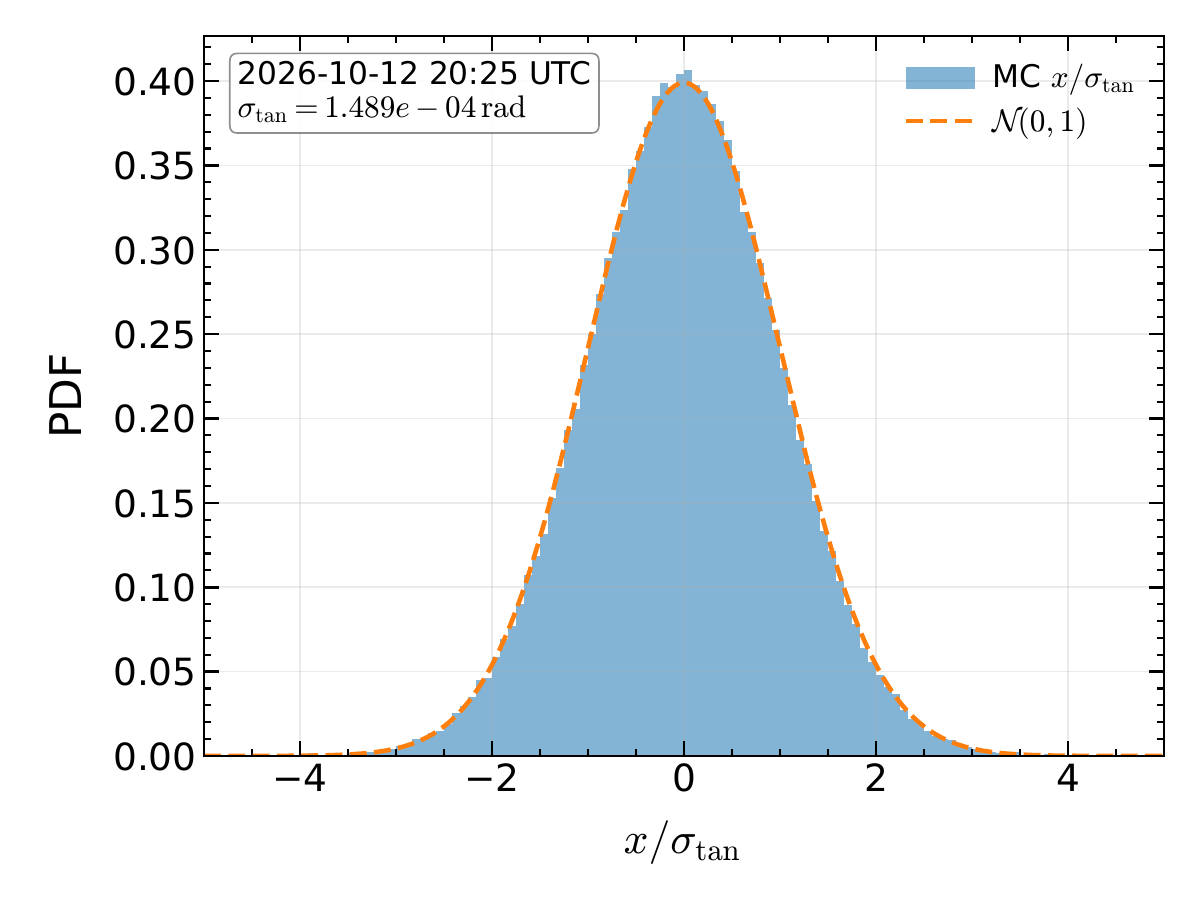}        
\caption{One-dimensional tangent-plane marginal distribution of \(x/\sigma_{\rm tan}\).}        
\label{fig:gaussian_cloud_signed_marginal}    
\end{subfigure}    
\hfill    
\begin{subfigure}[t]{0.48\textwidth}        
\centering        
\includegraphics[width=\textwidth]{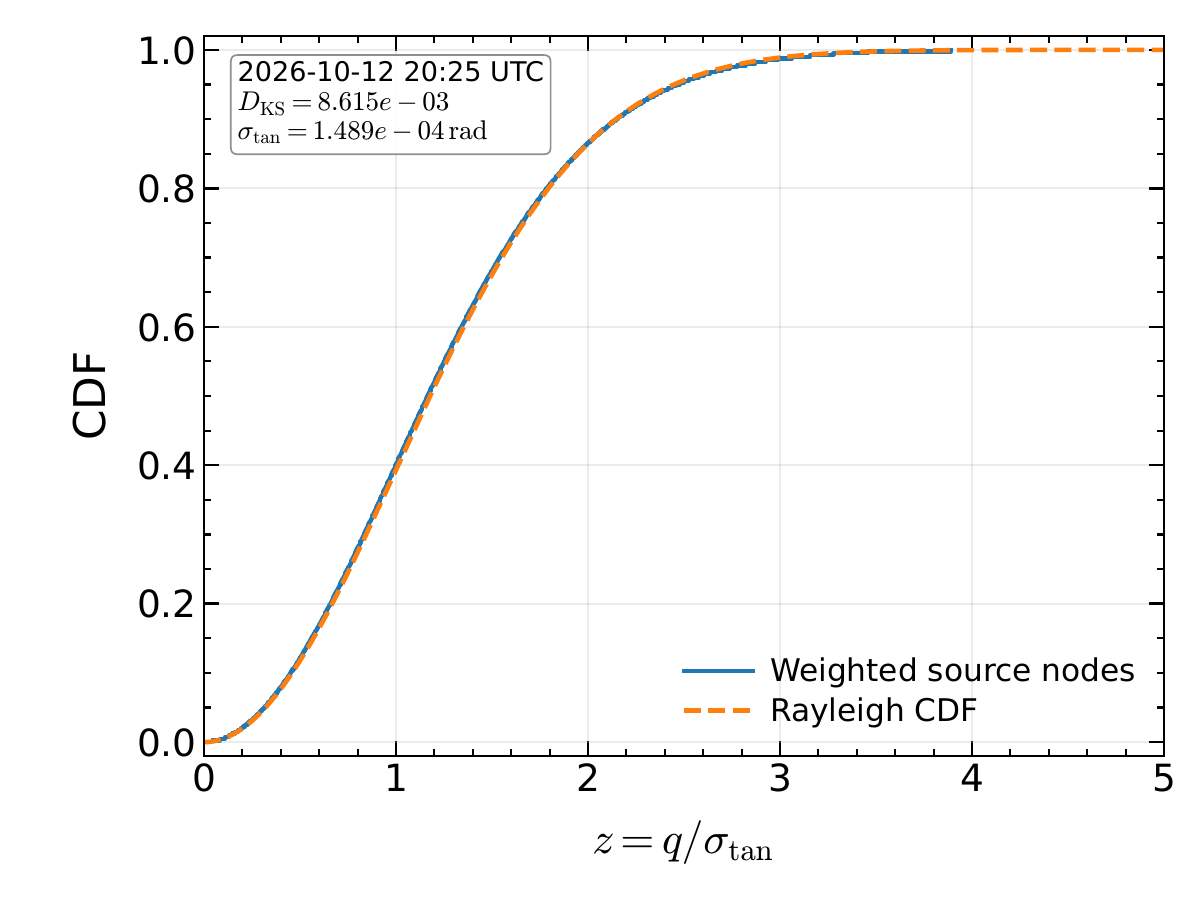}        
\caption{Radial cumulative distribution of \(z = q/\sigma_{\rm tan}\).}        
\label{fig:gaussian_cloud_radial_cdf}    
\end{subfigure}    
\caption{    
Numerical validation of the isotropic Gaussian-cloud approximation for a representative nighttime sample. Panel (a) compares the empirical distribution of the signed one-dimensional tangent-plane marginal \(x/\sigma_{\rm tan}\) against the standard normal distribution. Panel (b) compares the empirical radial CDF of \(z = q/\sigma_{\rm tan}\) against the theoretical Rayleigh CDF predicted by a two-dimensional isotropic Gaussian cloud.    
}    
\label{fig:gaussian_cloud_validation}
\end{figure}

Figure~\ref{fig:gaussian_cloud_validation} illustrates this same conclusion for a representative nighttime sample. The one-dimensional marginal distribution projected onto the tangent plane closely approximates $\mathcal{N}(0,1)$, and the radial cumulative distribution function (CDF) aligns closely with the theoretical Rayleigh prediction. Consequently, we adopt the isotropic Gaussian cloud as a low-dimensional summary to define the clustering distance. Nevertheless, the final finite-source reweighting is still carried out using ring-resolved source–node information—rather than substituting the physical source distribution with an exact Gaussian model.

\renewcommand{\thesubsection}{B.\arabic{subsection}}
\renewcommand{\thefigure}{B.\arabic{figure}}
\renewcommand{\theequation}{B.\arabic{equation}}
\section{Numerical Convergence Tests}
\label{app:convergence}

The target numerical precision for this work is at the $\mathcal{O}(10^{-4})$ level for probability-level observables. As a conservative convergence diagnostic, we employ maximum-norm differences over the computational energy grid; percentile-based, mean, and spectrum-integrated differences are typically smaller. The production settings are as follows: $N_\rho = 400$ and $\epsilon_{\rm pred} = 5 \times 10^{-6}$ on the solar side; $N_x = 8000$ layers for Strang-split propagation on the Earth side; and $K = 10^4$ representative nighttime trajectories for $k$-medoids path compression.

We adopt the production-averaged mass-eigenstate transition probabilities $P_{ei}^{\odot}$ as the solar-side diagnostic and the final nighttime electron-neutrino survival probability $P_{ee}^{N}$ as the Earth-side diagnostic. Unless otherwise specified, all results are benchmarked against a higher-resolution reference calculation. The subsequent validation tests confirm that the chosen settings control the relevant probability-level quantities to within a few parts in $10^{-4}$ or better.

\subsection{Solar-side convergence}
\label{app:solar-convergence}
\begin{figure}[!htbp]    
\centering
    \begin{subfigure}[t]{0.48\linewidth}        
\centering        
\includegraphics[width=\linewidth]        
{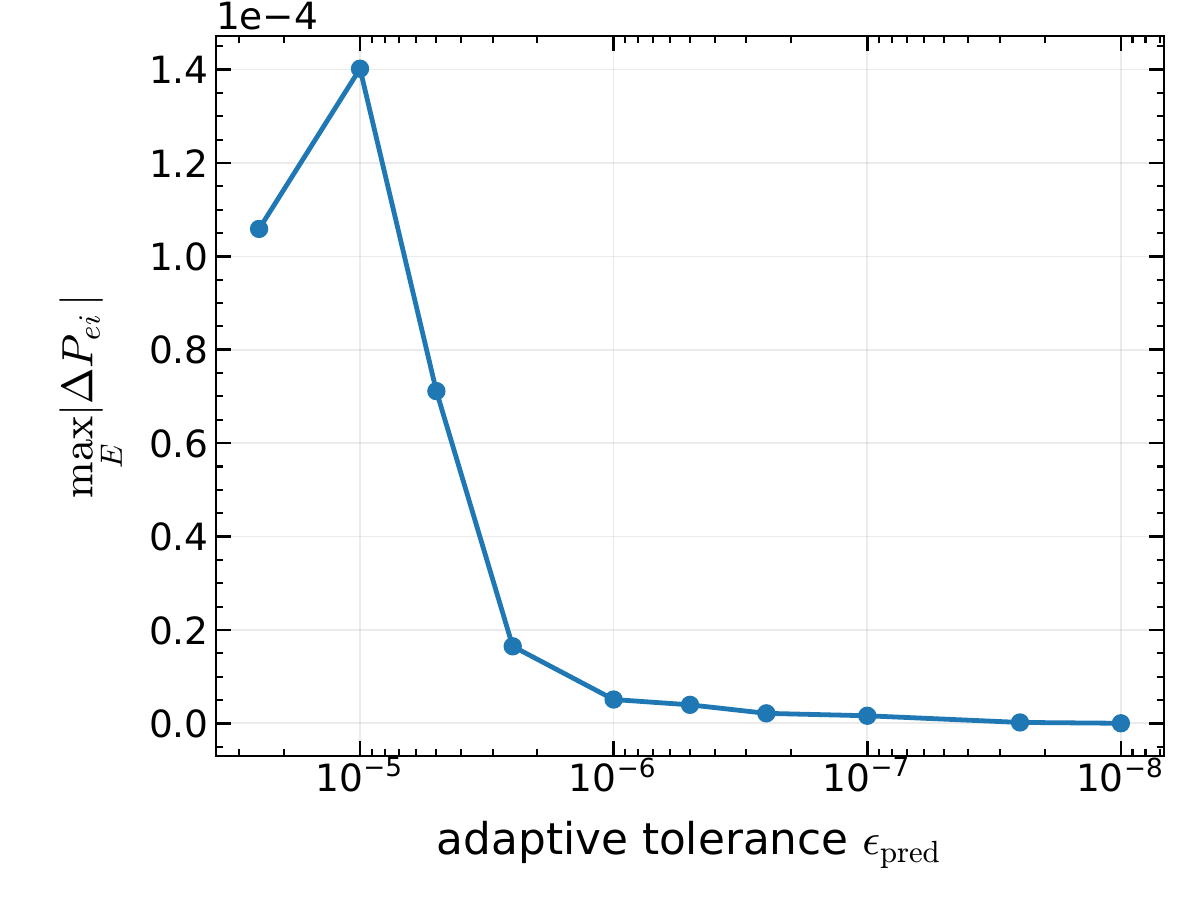}        
\caption{        
Convergence with respect to the adaptive propagation tolerance        
$\epsilon_{\rm pred}$. The result obtained with        
$\epsilon_{\rm pred}=10^{-8}$ serves as the reference solution.        
}        
\label{fig:solar-pei-tol-convergence}    
\end{subfigure}    
\hfill    
\begin{subfigure}[t]{0.48\linewidth}        
\centering        
\includegraphics[width=\linewidth]        
{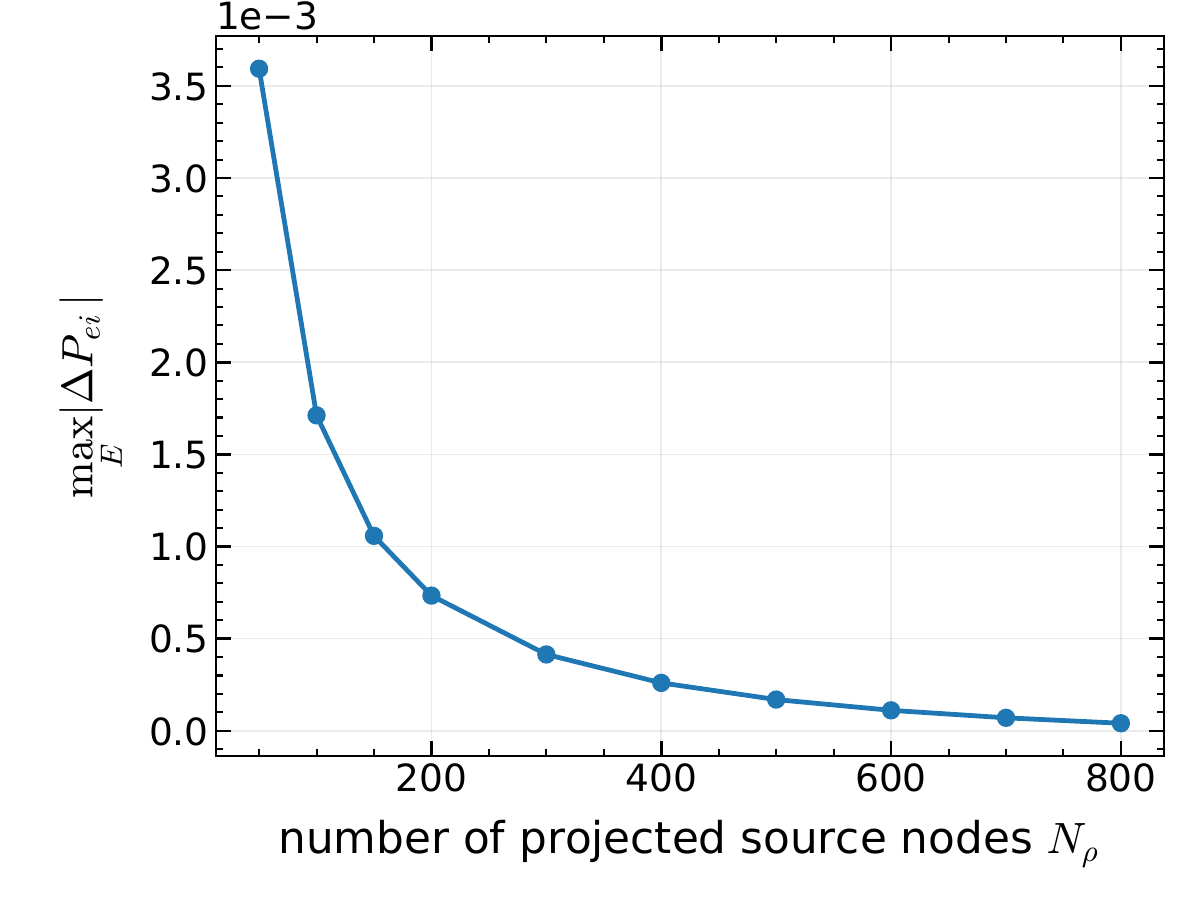}        
\caption{        
Convergence with respect to the discretization $N_\rho$ of the projected ${}^{8}\mathrm{B}$ neutrino source. The result for $N_\rho=1000$ is taken as the reference, with        
$\epsilon_{\rm pred}=5\times10^{-6}$ held fixed.        
}        
\label{fig:solar-source-disc-pei-convergence}    
\end{subfigure}
    \caption{    
Numerical convergence of the solar-side production-averaged    
mass-eigenstate transition probabilities $P_{ei}^{\odot}(E)$. In both panels, the plotted quantity is the maximum absolute deviation over the computational energy grid and across all three mass eigenstates:    
$\max_{E,i}|\Delta P_{ei}^{\odot}(E)|$.    
Panel~(a) assesses the sensitivity to the adaptive propagation tolerance, while    
panel~(b) evaluates the dependence on the projected source discretization.    
}    
\label{fig:solar-side-convergence}
\end{figure}
As shown in Fig.~\ref{fig:solar-side-convergence}, we examine the numerical convergence with respect to two independent settings: (i) the adaptive propagation tolerance and (ii) the discretization of the projected ${}^{8}\mathrm{B}$ neutrino production source. For the adaptive-tolerance test, we consider
\begin{align}
\epsilon_{\mathrm{pred}}
\in
\{
2.5\times10^{-5},
1\times10^{-5},
5\times10^{-6},
2.5\times10^{-6},
1\times10^{-6},\nonumber\\
5\times10^{-7},
2.5\times10^{-7},
1\times10^{-7},
2.5\times10^{-8},
1\times10^{-8}
\},\nonumber
\end{align}
with $\epsilon_{\mathrm{pred}}^{\mathrm{ref}}=1\times10^{-8}$ taken as the reference solution. For the source-discretization test, we use
\[
N_\rho
\in
\left\{
50,100,150,200,300,400,500,600,700,800,1000
\right\},
\]
with $N_\rho^{\mathrm{ref}}=1000$ taken as the reference, while fixing $\epsilon_{\mathrm{pred}}=5\times10^{-6}$.

For both tests, we report the maximum absolute deviation over the computational energy grid and across all three mass eigenstates,
\[
\max_{E,i}
\left|
\Delta P_{ei}^{\odot}(E)
\right|.
\]
This maximum-norm diagnostic provides a conservative measure of the numerical error; percentile-based or energy-averaged deviations are generally smaller.

The adaptive-tolerance test isolates the convergence behavior of the adaptive spatial mesh employed in solar propagation, whereas the source-discretization test quantifies the accuracy of the quadrature rule used to average over the extended production region. Both tests confirm that the solar propagation and production-averaging procedures are numerically stable and converge to the precision required in this study.

\subsection{Earth-side Strang-splitting solver and path-clustering convergence}
\label{sec:earth-side-convergence}
We assess two independent numerical approximations entering the Earth-side calculation: (i) the equal-length Strang discretization of each Earth-crossing trajectory and (ii) the compression of the nighttime trajectory ensemble using representative paths.

For the Earth-layer convergence test, we vary the number of equal-length Strang layers, \(N_x\), while keeping the solar-side output, geometric sampling, and representative Earth trajectories fixed. We consider
\[
N_x
\in
\left\{
500, 1000, 2000,3000, 4000,6000, 8000,10000,12000,15000,18000,20000
\right\},
\]
where \(N_x^{\mathrm{ref}}=20000\) is adopted as the reference baseline. The corresponding convergence diagnostic is
\begin{equation}
\max_E
\left|
\Delta P_{ee}^{N}(E;N_x)
\right|.
\end{equation}
For the trajectory-compression test, each value of \(K\) is evaluated by recomputing the full clustering pipeline, including the selection of medoid trajectories, the construction of their three-dimensional Earth density profiles, and the final solar--vacuum--Earth contraction.
The reference result retains all \(262\,167\) nighttime trajectories without clustering. We consider 
\[
K
\in
\left\{
500, 1000, 2000, 4000, 6000, 8000, 10000,12000, 15000,18000
\right\},
\]
and define
\begin{equation}
\Delta P_{ee}^{N}(E;K)
=
P_{ee}^{N}(E;K)
-
P_{ee}^{N}(E;\mathrm{direct}),
\end{equation}
and report
\begin{equation}
\max_E
\left|
\Delta P_{ee}^{N}(E;K)
\right|.
\end{equation}
\begin{figure}[!htbp]
    \centering

    \begin{subfigure}[t]{0.48\textwidth}
        \vspace{0pt}
        \centering
        \includegraphics[width=\linewidth]
        {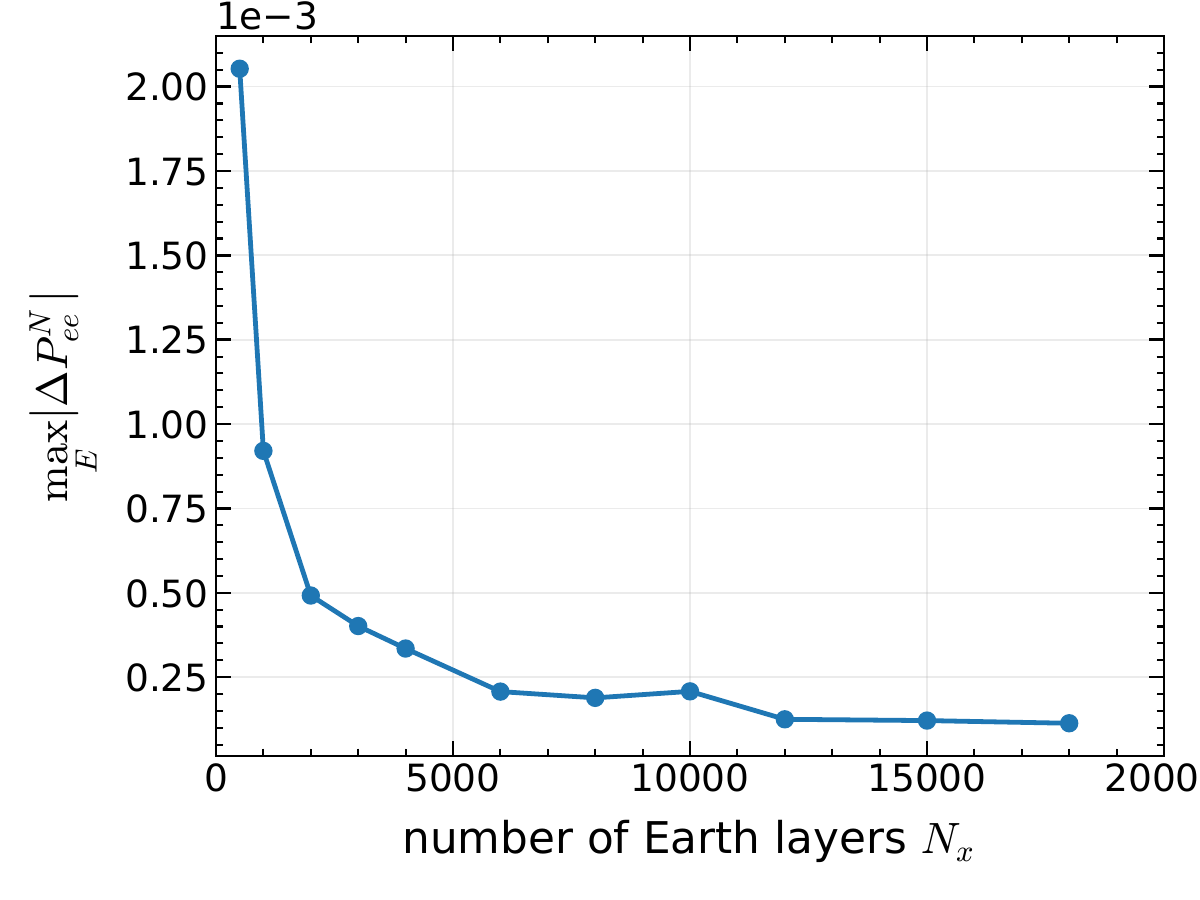}
        \caption{
        Convergence with respect to the number of equal-length
        Earth layers \(N_x\). The calculation with
        \(N_x^{\mathrm{ref}}=20000\) serves as the reference.
        }
        \label{fig:earth-layer-convergence}
    \end{subfigure}
    \hfill
    \begin{subfigure}[t]{0.48\textwidth}
        \vspace{0pt}
        \centering
        \includegraphics[width=\linewidth]
        {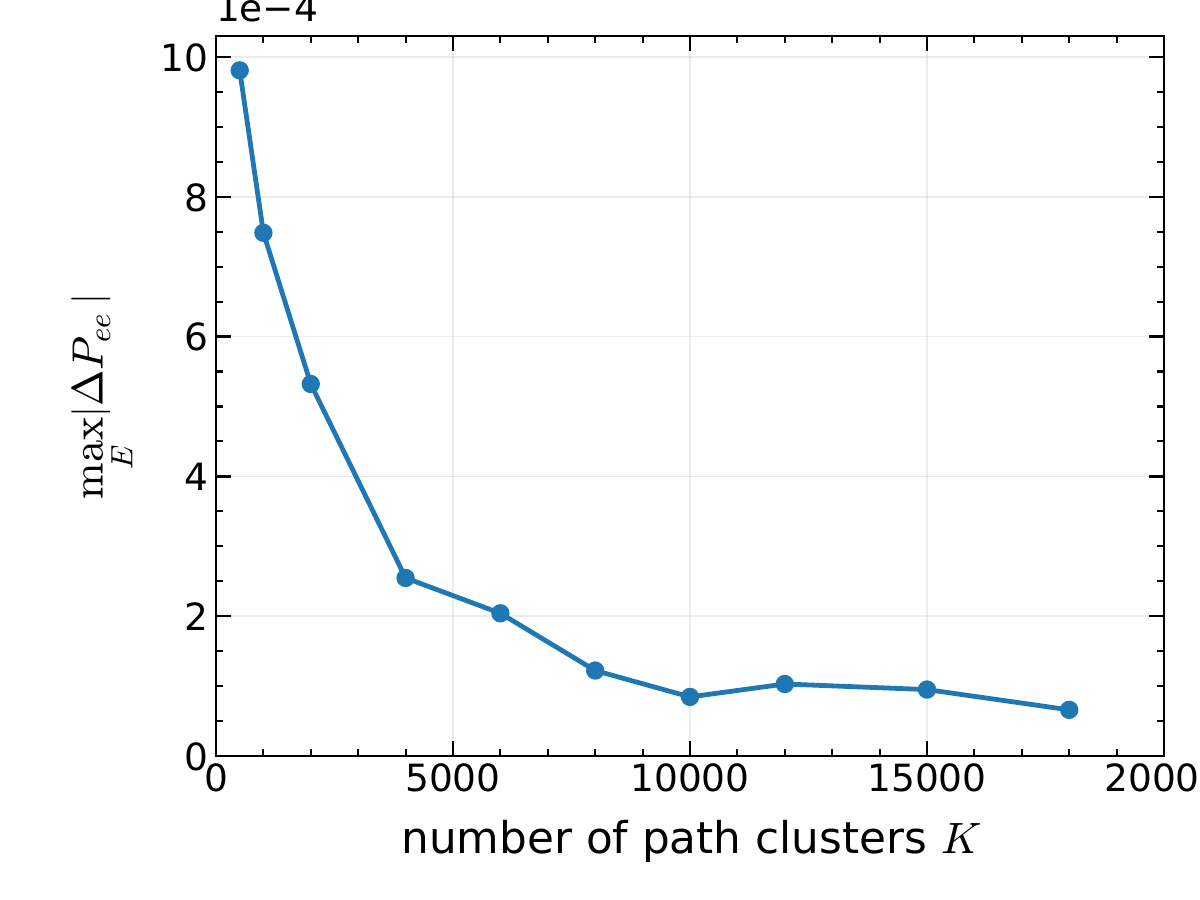}
        \caption{
        Convergence with respect to the number of representative
        nighttime trajectories \(K\), relative to the unclustered
        direct calculation.
        }
        \label{fig:path-clustering-convergence}
    \end{subfigure}

    \caption{
    Numerical convergence of the Earth-side calculation.
    Panel~\subref{fig:earth-layer-convergence} shows the convergence
    of the nighttime electron-neutrino survival probability with
    respect to the number of equal-length Strang layers.
    Panel~\subref{fig:path-clustering-convergence} shows the
    convergence of the representative-path compression relative
    to the calculation retaining all nighttime trajectories.
    }
    \label{fig:earth-side-convergence}
\end{figure}

At the adopted production settings, the maximum pointwise deviations are approximately \(1.9\times10^{-4}\) for \(N_x=8000\), and \(8.5\times10^{-5}\) for \(K=10^4\), respectively. The maximum-norm diagnostic provides a conservative measure of the numerical error; percentile-based and energy-averaged deviations are generally smaller. These results are consistent with the few-parts-in-\(10^{-4}\) numerical-accuracy target adopted in this work and remain small compared with the current parametric uncertainties.

\renewcommand{\thesubsection}{C.\arabic{subsection}}
\renewcommand{\thefigure}{C.\arabic{figure}}
\renewcommand{\thetable}{C.\arabic{table}}
\section{Cross-code validation and computational performance}
\label{app:speed}
We also provide an independent validation of the Earth-side with \textsc{nuSQuIDS}~\cite{nuSQuIDS} for three representative Earth-crossing baselines(obtained from 1D PREM). The representative wall-clock times and probability-level differences are summarized in Table~\ref{tab:nusquids_validation}.

\begin{table}[!htbp]
    \centering
    \caption{
    Cross-code validation of the Earth-side Strang-splitting solver against
    \textsc{nuSQuIDS}. The timings correspond to a single trajectory evaluated
    on the same energy grid. The speed-up is defined as
    $t_{\textsc{nuSQuIDS}}/t_{\rm Strang}$.
    }
    \label{tab:nusquids_validation}
    \begin{tabular}{ccccccc}
        \hline
        $L$ [km]
        & Core crossing
        & $t_{\rm Strang}$ [s]
        & $t_{\textsc{nuSQuIDS}}$ [s]
        & Speed-up
        & $\max |\Delta P|$
        & RMS$(\Delta P)$ \\
        \hline
        1000
        & No
        & 0.02299
        & 0.7995
        & 34.8
        & $2.36\times10^{-5}$
        & $2.09\times10^{-6}$ \\

        6000
        & No
        & 0.1385
        & 5.527
        & 39.9
        & $1.22\times10^{-5}$
        & $1.28\times10^{-6}$ \\

        11000
        & Yes
        & 0.2621
        & 10.97
        & 41.9
        & $3.04\times10^{-5}$
        & $8.63\times10^{-6}$ \\
        \hline
    \end{tabular}
\end{table}
All single-trajectory benchmarks were performed on an AMD Ryzen Threadripper 7980X using 64 OpenMP threads. For all three representative trajectories, the maximum absolute probability difference remains below $3.1\times10^{-5}$, while the root-mean-square difference remains below $8.7\times10^{-6}$. The Strang-splitting implementation reduces the wall-clock propagation time by a factor of approximately $35$--$42$ for these single-trajectory tests. We emphasize that \textsc{nuSQuIDS} is a general-purpose neutrino-propagation framework supporting a substantially broader range of physical processes, whereas the present solver is specialized for coherent three-flavor propagation in prescribed matter profiles. The comparison is therefore
intended primarily as an independent cross-code validation and as a representative characterization of the computational cost for the specific workload considered here.

We also test a batch consisting of 100 Earth-crossing trajectories, 2000 neutrino energies, and a total of $1.15\times10^{6}$ matter layers; the corresponding wall-clock times are approximately 17.0~s for the 64-thread \textsc{nuSQuIDS} calculation and 0.221~s for the 64-thread CPU Strang implementation, corresponding to a speed-up of approximately 77. The GPU RawKernel implementation evaluates the same workload in approximately 0.0759~s on an NVIDIA RTX~5090 and 0.0171~s on an NVIDIA A100.

For the default production configuration adopted in this work, we use $N_x=8000$ Earth-side layers, $K=10\,000$ representative nighttime trajectories, an energy grid of $N_E=2000$ points, and a solar-source discretization of $N_\rho=400$, with an adaptive-mesh tolerance of $\epsilon_{\rm pred}=5\times10^{-6}$. 
For the solar-side propagation described in Sec.~\ref{sec:solar_msw}, the full three-flavor propagation kernel requires approximately 2~s on an NVIDIA A100 after kernel compilation and warm-up. Following the same compilation and warm-up procedure, the Earth-side Strang-splitting kernel requires approximately 0.9~s.

For a single detector site, the complete end-to-end setup and propagation workflow requires approximately 5~min on on the computing platform used in this work(excluding the Monte Carlo). This wall-clock time includes the construction of the time-dependent Sun--detector geometry, the solar adaptive mesh, the representative-path clustering, the three-dimensional Earth matter-potential grids, and the subsequent solar- and Earth-side propagation. Most of this total time is associated with one-time geometric and preprocessing operations rather than with the propagation kernels themselves.
\bibliography{mybibfile}

\end{document}